\documentclass[preprint,preprintnumbers,amsmath,amssymb,prd,tightenlines
]{revtex4-1}
\usepackage{graphicx}
\begin{document}
\setlength{\voffset}{.5cm} 
\preprint{RIKEN-MP-08}
\title{Static interactions and  stability of matter in Rindler space}
\newpage
\author{F. Lenz $^{{\rm a,c}}$}    \email{flenz@theorie3.physik.uni-erlangen.de} 
\author{K. Ohta $^{{\rm b}}$}      \email{ohta@nt1.c.u-tokyo.ac.jp}
\author{K. Yazaki $^{{\rm c,d}}$}  \email{yazaki@phys.s.u-tokyo.ac.jp}
\affiliation{$^{{\rm a}}$ Institute for Theoretical Physics III \\
University of Erlangen-N\"urnberg \\
Staudtstrasse 7, 91058 Erlangen, Germany\\ \\
$^{{\rm b}}$ Institute of Physics \\ University of Tokyo\\ Komaba, Tokyo 153, Japan  \\ \\
$^{{\rm c}}$ Hashimoto Mathematical Physics Laboratory\\ Nishina Center, RIKEN\\
Wako, Saitama 351-0198, Japan \\ \\
$^{{\rm d}}$ Yukawa Institute for Theoretical Physics, \\ Kyoto University\\ Kyoto 606-8502, Japan \\}
\date{December 15, 2010} 
\begin{abstract}
Dynamical issues associated with quantum fields in Rindler space are addressed in a study of the interaction between two  sources at rest  generated by the exchange of scalar particles, photons and gravitons. These static interaction energies  in Rindler space are shown to be  scale invariant, complex quantities.  The imaginary part will be seen to have its quantum mechanical origin in the presence of an infinity of zero modes in uniformly accelerated frames which in turn are related to the radiation observed in inertial frames. The impact of a uniform  acceleration on the stability of matter and the properties of particles is discussed and  estimates are presented of the instability of hydrogen atoms when approaching the horizon. 
\end{abstract} 
\maketitle
\section{Introduction}
The study of quantum fields in Rindler space has played an important role in developing our understanding of quantum fields in non-trivial space-times. The importance of these studies derives to a large extent from the existence of a horizon in Rindler space. With its close connection to Minkowski space via the  known local relation between fields in uniformly accelerated and inertial frames, Rindler space provides the simplest possible context to investigate kinematics and dynamics of quantum fields close to a horizon.

The kinematics of non-interacting quantum fields in Rindler space \cite{FULL73},\,\cite{BOUL75},\,\cite{DAVI75} and their relation to fields in Minkowski space  together with the interpretation in terms of quantum fields at finite temperature  \cite{UNRU76},\,\cite{SCCD81} are  well understood. The relation between acceleration and finite temperature remains an important element of the thermodynamics of black holes. Although formally established, the relation between fields in Rindler and Minkowski spaces has posed some intriguing interpretational problems. Here we mention in particular the issue of compatibility of the radiation generated by a uniformly accelerated charge  observed in Minkowski space and the apparent absence of radiation in the coaccelerated frame. This problem was finally resolved by the realization that the counterpart of  Minkowski space radiation is the emission of zero energy photons in Rindler space \cite{HIMS921},  \cite{HIMS922},   \cite{REWE94}.  
Dynamical issues of quantum fields have received less attention (cf.\,\cite{CRHM07}). Related to topics to be discussed in the following are  the  study  of the relation between interacting quantum fields in Minkowski and Rindler spaces within the path-integral formalism \cite{UNRU84}, the calculation of level shifts in accelerated hydrogen atoms \cite{PASS98} and the investigations of the decay of protons \cite{MULL97},\,\cite{VAMA00}. 

Dynamical issues of quantum fields in Rindler space are in the center of our work.  Subject are  static forces  and their properties relevant for the structure of matter  in Rindler space. We will study the forces acting between static scalar sources, electric charges and massive sources generated by exchange of massless scalar particles,  photons and gravitons respectively. The relevant quantities involved are the corresponding static Rindler space propagators. While the time dependent propagators in Rindler and  Minkowski spaces are trivially related  to each other by a coordinate transformation with corresponding mixing of the components, this is not the case   for  the static propagators obtained by integration over the corresponding time.  The different properties of the static propagators reflect the significant differences of Rindler and Minkowski space Hamiltonians. Though it was realized from the beginning that the interpretation of Rindler and Minkowski particles (cf.\,e.g.\,\cite{FULL73}) is different, the differences in the corresponding Hamiltonians have not received sufficient attention.  (The identity of the Rindler and Minkowski space Hamiltonians for the frequently used special case of a massless field in two dimensional space-time  may have obscured the issue.) As shown in \cite{LOY08}, the Rindler space Hamiltonian exhibits symmetries which give rise to a  highly degenerate spectrum. In particular this degeneracy is also present in the  zero energy sector. This 2-dimensional sector of zero modes is the image of the photons generated by the charges uniformly accelerated in Minkowski space. It will be shown to give rise to an unexpected quantum mechanical contribution to the force acting between two charges. Also the  classical Coulomb-like contribution to the force will be shown to significantly deviate from the electrostatic force in Minkowski space. The  electrostatic force will be determined  via  Wilson loops  and  Polyakov loop correlation functions. This method  will enable us to separate the contribution  of the quantum mechanical  transverse photons from that of the classical longitudinal field. It will be  the method of choice  if one attempts to determine the static force   in  simulations of non-abelian gauge theories on a Rindler space lattice.

The modifications in the interaction of two charges at rest in Rindler space raise naturally the question of the stability of atomic systems in Rindler space. We shall investigate this issue for an ensemble of hydrogen atoms and show within a non-relativistic reduction that indeed only metastable states exist. We will estimate the probability of ionization as a function of the distance to the horizon and present arguments concerning the stability of other forms of matter.     
\section{Propagators and interactions of scalar particles  in Rindler space}
\subsection{The quantized scalar field in Rindler space}
The Rindler space  metric \cite{RIND01}
\begin{equation}
ds^{2} = e^{2 a \xi} (d \tau^{2} - d \xi^{2})- d{\bf x}^2_{\perp} \,,              
\label{rin}
\end{equation}
is the (Minkowski) metric seen by a uniformly accelerated observer (acceleration $a$ in $x^1$-direction). 
Rindler space ($\tau, \xi$)    and Minkowski space ($t, x^1$) coordinates  are related by
\begin{equation}
t (\tau, \xi) = \frac{1}{a} \; e^{a \xi} \sinh a \tau\,, \quad 
x^1 (\tau, \xi) = \frac{1}{a} \; e^{a \xi} \cosh a \tau \, .          
\label{cd3}
\end{equation}
The range of $\tau,\,\xi $ is 
\begin{equation}
- \infty \; \le \; \tau\,,\xi \; \le \; \infty \,,                               \label{rau2}
\end{equation}
while    the preimage  of the Rindler space covers only part of Minkowski space, the right ``Rindler wedge'',
\begin{equation}
R_+ = \big\{x^\mu\big|\,|t|\le x^1 \big\}\,.                          
\label{rau3}
\end{equation}
The restriction of the preimage of the  Rindler space to the right Rindler wedge gives rise to a horizon, the boundary  $t=\pm x^1,\,\,\text{i.e.}\,\,  \xi=-\infty$. With this property the Rindler metric can be identified with other static metrics in the near horizon limit. In particular this is the case for the Schwarzschild metric which can  be approximated in the limit that the distance from the horizon is small in comparison to the Schwarzschild radius  and if the spherical Schwarzschild horizon is replaced by a tangential plane. 
  
We start our study of propagators and interactions with a discussion of a scalar field in Rindler space (cf.\,for instance\,\cite{FULL73}, \cite{CRHM07}. We will use the notation of \cite{LOY08}).  The action of a free massive  scalar field is given by
\begin{equation}
S = \frac{1}{2} \int d \tau \; d \xi \; d^{\,2}x_{\perp} \big \{ \partial_{\tau} \phi )^{2}
- (\partial_{\xi} \phi )^{2} - (m^{2} \phi^{2} + (\boldsymbol{\partial}_{\perp} \phi)^{2} ) \;
 e^{2 a \xi} \big \} \, .                                                           \label{ac}
\end{equation}
The wave equation  
\begin{equation}
\Big [ \partial^{2}_{\tau} - \Delta_s + m^{2} e^{2 a \xi} \Big ] \phi = 0   \, ,                              
\label{EQM1}
\end{equation}
with the Laplacian
\begin{equation}
\label{scala}
\Delta_s = \partial^{2}_{\xi} +e^{2 a \xi}\partial_{\perp}^2\, , 
\end{equation}
is solved by the  normal modes (in terms of the McDonald functions)
\begin{equation}
\phi_{\omega,{\bf k}_\perp}(\xi,{\bf x_\perp}) =\frac{1}{\pi} \; \sqrt{2 \frac{\omega}{a} \sinh  \pi \frac{\omega}{a} }  \;  K_{i  \frac{\omega}{a}} \Big(\frac{1}{a}\sqrt{m^2+{\bf k}^2_\perp}e^{a\xi}\Big) \, e^{i {\bf k} _{\perp} {\bf x}_{\perp}}  \, ,
\label{nomde}
\end{equation} 
which form a complete, orthonormal set of functions. 
The normal mode expansion of the scalar field $\phi$ reads 
\begin{eqnarray}
\phi (\tau, \xi, {\bf x}_{\perp}) =  \int \frac{d \omega}{\sqrt{2\omega}} \frac{d^{2}k_{\perp}}{2\pi}
\Big(a (\omega, {\bf k}_{\perp}) \phi_{\omega,{\bf k}_\perp}(\xi,{\bf x_\perp})e^{-i\omega \tau}+ a^{\dagger}(\omega, {\bf k}_{\perp}) \phi^\star_{\omega,{\bf k}_\perp}(\xi,{\bf x_\perp}) e^{i\omega \tau}\Big)
\label{nomosc}
\end{eqnarray}
with the creation  and annihilation operators $a^{(\dagger)}$. The stationary states   associated with the Hamiltonian ($\pi = \partial_{\tau} \phi) \,$
\begin{eqnarray}
H_R &=&  \frac{1}{2}\int d\xi d^2x_\perp \big \{  \pi^{2} + (\partial_{\xi}\phi)^{2}
+\, e^{2 a \xi} \big(m^{2} \phi^{2} + (\boldsymbol{\partial}_{\perp}\phi)^{2}\big) \big \}\nonumber\\ &=&\int d^{2}k_{\perp}\int^{\infty}_{0} d \omega\, \omega  a^{\dagger} (\omega, {\bf k}_{\perp}) a (\omega, {\bf k}_{\perp})
 \,  ,
\label{hamsca}
\end{eqnarray}
including the lowest energy ($\omega=0$) state, 
are  degenerate with respect to the value of the transverse momentum ${\bf k}_\perp$. This degeneracy has its origin in the appearance of   the transverse momentum (in combination with the mass)   as a coupling constant of the ``inertial potential'' $e^{ 2a\xi}$ in the  Hamiltonian.

In the Rindler wedge (\ref{rau3}), the field operator $\phi$ can be represented in terms of  plane waves in Minkowski space or in terms the Rindler space normal modes (\ref{nomosc}) resulting in the representation of the Rindler creation and annihilation operators in terms of the corresponding Minkowski space operators. This relation yields the important result for the expectation value of the Rindler number operator in the Minkowski ground state $|0_M\rangle$
\begin{equation}
\langle 0_M | a^{\dagger} (\omega, {\bf k}_{\perp})  a (\omega^{\prime}, {\bf k}_{\perp}^{\prime}) | 0_M \rangle
=\frac{1}{e^{2 \pi \frac{\omega}{a}} - 1} \delta (\omega - \omega^{\prime})\delta({\bf k}_{\perp}-{\bf k}^{\prime}_{\perp}) \,         
\label{urt2} 
\end{equation}     
which exhibits a thermal distribution with temperature 
\begin{equation}
T=\frac{1}{\beta}=\frac{a}{2\pi}\,.
\label{temp}
\end{equation} 
It is important to realize  that the infinite degeneracy of the Rindler eigenstates leads, at long wavelengths, effectively to 1-dimensional thermal distributions with weight  
$d\omega/(e^{2 \pi \frac{\omega}{a}} - 1)$\,. 
\subsection{The scalar Rindler space propagator }  \label{sec:scpr}
The propagators  in Minkowski and Rindler spaces are  related to each other by the coordinate transformation (\ref{cd3}). Written in terms of Rindler and Minkowski coordinates respectively, the propagator of a massless field is given by (cf.\,\cite{TRDA77},\,\cite{DOWK78})
\begin{eqnarray}
&&D(\tau-\tau^\prime, \xi, \xi',{\bf x}_{\perp}-{\bf x}_{\perp}^\prime)
= i \langle 0_M\big| T\big[\phi(\tau, \xi, {\bf x}_{\perp}) \phi(\tau^\prime,\xi',{\bf x}_{\perp}^\prime)\big]\big|0_M\rangle = D(x-x^\prime)\nonumber\\ &&=\frac{1}{4i\pi^2\big((x-x^\prime)^2 -i\delta\big)}= \frac{a^2e^{-a(\xi+\xi^\prime)}}{8i\pi^2} \frac{1}{\cosh a(\tau-\tau^\prime)  -\cosh \eta -i\delta}\,, 
\label{ripo3}
\end{eqnarray} 
with the notations 
\begin{equation}
D(\tau-\tau^\prime, \xi, \xi',{\bf x}_{\perp}-{\bf x}_{\perp}^\prime)
=D(x(\tau,\xi,{\bf x}_\perp)-x(\tau^\prime,\xi^\prime,{\bf x^\prime}_\perp))
\label{DD}
\end{equation}
and
\begin{eqnarray}
\cosh\eta(\xi,\xi^\prime,{\bf x}_\perp-{\bf x}_{\perp}^\prime) =  1+ \frac{\big(e^{a\xi}-e^{a\xi^\prime}\big)^2+ a^2\big({\bf x}_\perp-{\bf x}_\perp^{\prime}\big)^2}{2 e^{a(\xi+\xi^\prime)}}=1+\sigma^2(\xi,\xi^\prime,{\bf x}_\perp-{\bf x}_{\perp}^\prime)\,.
\label{ueta}
\end{eqnarray}
As we will see, it is $\eta$ (or equivalently $\sigma$) which  determines the interaction energies rather than  the proper distance in Rindler space
\begin{equation}
d(\xi,\xi^\prime,{\bf x}_\perp-{\bf x}_{\perp}^\prime)=\sqrt{\frac{1}{a^2}\big(e^{a\xi}-e^{a\xi^\prime}\big)^2+\big({\bf x}_\perp-{\bf x}_{\perp}^\prime\big)^2}\,.
\label{prdi}
\end{equation}
 The quantity $\eta$  actually can be viewed as the proper distance in the 4-dimensional AdS space. The appearance of this geometry (of a three-hyperboloid) has been noted \cite{SCCD81} in the context of the unusual density of states in the photon energy momentum tensor in Rindler space. The quantity $\eta$ or equivalently $\sigma$  exhibits the remarkable invariance under the transformation 
\begin{equation}
\label{symm1}
\xi,\,\xi^\prime \to \xi+\xi_0,\,\xi^\prime+\xi_0\,,\quad {\bf x}_\perp,\, {\bf x}_\perp^{\prime}\to e^{a\xi_0}{\bf x}_\perp,\, e^{a\xi_0}{\bf x}_\perp^{\prime}\,.
\end{equation}
The coordinate transformation (\ref{symm1}) together with a rescaling of the fields $\phi(\tau,\xi,{\bf x}_\perp)\to e^{a\xi_0}\phi(\tau,\xi,{\bf x}_\perp)$ leaves, for $m=0$,  the Hamiltonian (\ref{hamsca}) invariant and gives rise to the degeneracy of the spectrum. The invariance under scale transformations can be generalized to massive particles \cite{LOY08}.

In   Rindler space coordinates the 2-point function (\ref{ripo3}) satisfies 
\begin{equation}
\big(\partial_\tau^2-\partial_\xi^2-e^{2a\xi}\partial_\perp^2 \big) D(\tau, \xi,\xi^\prime,{\bf x}_\perp) = \delta(\tau)\delta(\xi-\xi^\prime) \delta({\bf x}_\perp)\,.
\label{ripo2}
\end{equation}
After a Wick rotation of the Rindler time, 
\begin{equation}
\tau\to\tau_E=-i\tau\,,
\label{wiro}
\end{equation}
the propagator is periodic in the imaginary time with periodicity $\beta$  (cf.\,Eq.\,(\ref{temp}))
\begin{equation}
 D_{\text{E}}(\tau_{\text{E}}, \xi, \xi',{\bf x}_{\perp})=
 \frac{a^2e^{-a(\xi+\xi^\prime)}}{8i\pi^2} \frac{1}{\cos a \tau_{\text{E}}  -\cosh \eta -i\delta}\,. 
\label{perpro}
\end{equation}
In the context of the electromagnetic field we will compute  observables in  both the  real and imaginary Rindler time formalisms.  
(For discussions of  ``Euclidean'' Rindler space propagators, in particular of their topological interpretation  cf.\,\cite{CHDU78},\,\cite{LINE95},\,\cite{SVZA08}.)

Without reference to the Minkowski space propagator,  Eq.\,(\ref{ripo3}) can be derived alternatively  via the normal mode decomposition (\ref{nomosc}) in Rindler space. With the help of  Eq.\,(\ref{urt2})  the result  
\begin{eqnarray}
D(\tau, \xi, \xi',{\bf x}_{\perp})
= D^{(R)}(\tau, \xi, \xi',{\bf x}_{\perp})+i\int_0^\infty \frac{d\Omega}{\Omega} \int \frac{d^2  k_{\perp}}{(2\pi)^2} 
\phi_{\Omega,{\bf k}_\perp}(\xi,{\bf x_\perp})\phi_{\Omega,{\bf k}_\perp}(\xi^\prime,{\bf 0_\perp}) 
\frac{\cos{\Omega \tau}}
{e^{\frac{2\pi \Omega}{a}} - 1}\nonumber\\
\label{Dur}
\end{eqnarray}
is obtained where the propagator defined with respect to  the Rindler space vacuum is given by
\begin{eqnarray}
&&\hspace{-1.3cm}D^{(R)}(\tau, \xi, \xi',{\bf x}_{\perp})
= i \langle 0_R\big| T\big[\phi(\tau, \xi, {\bf x}_{\perp}) \phi(0, \xi', {\bf 0}_{\perp})\big]\big|0_R\rangle\nonumber\\&& \hspace{-1.3cm}= i \int_0^\infty \frac{d\Omega}{2\Omega} \int \frac{d^2  k_{\perp}}{(2\pi)^2} 
\phi_{\Omega,{\bf k}_\perp}(\xi,{\bf x_\perp})\phi_{\Omega,{\bf k}_\perp}(\xi^\prime,{\bf 0_\perp}) 
e^{-i\Omega |\tau|}= \frac{a^2e^{-a(\xi+\xi^\prime)}\eta}{4i\pi^2\,\sinh \eta} \frac{1}{a^2 \tau^2  -\eta^2 -i\delta}\,.
\label{d22}
\end{eqnarray}
 (For an interpretation of   the difference between the two propagators $D-D^{(R)}$ in terms of image charges located in the left Rindler wedge $|t|\le -x^1$, 
cf.\,\cite{CARA76}.) Obviously, the propagator $D^{(R)}$ depends parametrically on the acceleration $a$. In order to formulate properly the relation between propagators in Minkowski and  Rindler spaces  one has to  define the Rindler space propagator for a  fixed value of the acceleration $a$ at finite temperature (given by $\beta$) 
\begin{equation}
D_{a,\beta}(\tau,\xi,\xi^\prime,{\bf x}_\perp) = i\frac{1}{\text{tr}\, e^{-\beta H_R(a)}} \text{tr}\,\Big\{ e^{-\beta H_R(a)} T\big[\phi(\tau, \xi, {\bf x}_{\perp}) \phi(0, \xi', {\bf 0}_{\perp})\big]\Big\}\,.  
\label{thpri}
\end{equation}
In  terms of the propagator $D_{a,\beta}$, the central result concerning the relation between Rindler and Minkowski spaces is the identity      
\begin{equation}
 i\langle 0_M\big| T\big[\phi(\tau, \xi, {\bf x}_{\perp}) \phi(0, \xi', {\bf 0}_{\perp})\big]\big|0_M
\rangle =   D_{a,\frac{2\pi}{a}}(\tau,\xi,\xi^\prime,{\bf x}_\perp)\,,
\label{ftpro}
\end{equation} 
i.e.,\,the Rindler space propagator defined with respect to the Minkowski ground state coincides with the Rindler space finite temperature propagator with the value of the temperature determined by the acceleration $a$ (cf.\,Eq.\,(\ref{temp})).  This identity makes also manifest that a change in the acceleration $a$ does not correspond to  a change in temperature of the accelerated system. The acceleration appears not only as temperature in the Boltzmann factor but also as a parameter in the Hamiltonian (\ref{hamsca}) of the accelerated system.  We will encounter observables which make explicit this twofold role of the acceleration.

The basic quantity in the following investigation of the properties and consequences of  interactions generated  by exchange of scalar particles and photons is the static propagator 
\begin{equation}
\tilde{D}(\omega, \xi, \xi',{\bf x}_{\perp}) = \int_{-\infty}^\infty d\tau e^{i\omega \tau}  D(\tau, \xi, \xi',{\bf x}_{\perp})\,.
\label{stapo}
\end{equation}  
For vanishing mass the $k_\perp$-integration in (\ref{Dur}) can be carried out in closed form (cf.\,\cite{RG65} and \cite{EMOT53})
\begin{equation}
\label{saripo2ab}
\tilde{D}(\omega, \xi, \xi', {\bf x}_{\perp})  = 
\frac{a}{4\pi e^{a(\xi+\xi^\prime)} \sinh{\eta}} 
\Big(e^{i\frac{|\omega| \eta}{a}} + 
\frac{2i \sin{\frac{|\omega| \eta}{a}}}{e^{\frac{2\pi |\omega|}{a}} - 1}\Big)\,.
\end{equation}
Alternatively, this result can be obtained by a contour integration  of Eq.\,(\ref{ripo3}) in the complex $\tau$ plane. The first term is due to  the pole infinitesimally close to the real axis $\tau=\eta+i\delta$ while the second, imaginary contribution is the result of the  summation of the residues of the poles at $\tau_n =\eta + 2in \pi,\; n\ge 1$. 
The  result for the static propagator and its decomposition  into the real non-thermal and the imaginary thermal contributions  read
\begin{equation}
\tilde{D}(0,\xi,\xi^\prime,{\bf x}_\perp)=\tilde{D}^{(R)}(0,\xi,\xi^\prime,{\bf x}_\perp)+\frac{i}{2\pi \beta}\frac{\eta}{ e^{a(\xi+\xi^\prime)}\,\sinh \eta}\big|_{\beta=\frac{2\pi}{a}}= \frac{a}{4\pi e^{a(\xi+\xi^\prime)}}\frac{1}{\sinh \eta}\Big[ 1+\frac{ i \eta}{\pi} \Big] \,.
\label{saripo2a}
\end{equation}
The  imaginary part of the static propagator $\tilde{D}$ arises since the propagator (\ref{Dur}) is defined with respect to the Minkowski rather than to the Rindler ground state.
\subsection{The interaction energy of scalar sources} \label{sec:ipsp}
Given the propagator, the interaction energy between two  scalar sources is obtained   by adding to   the action (\ref{ac}) the scalar particle-source vertex
\begin{equation}
S_{\text{int}} = -\int dv^0 H_{\text{int}}=\kappa_0\int d^4 v \sqrt{|g(v)|}\,\rho(v)\,\phi(v)\,\,,
\label{coac}
\end{equation}
where  $v$ stands for either Minkowski  ($x$) or  Rindler coordinates ($\tau,\xi,{\bf x}_\perp$).
The effective action (the generating functional of connected diagrams) associated with the source   is given by 
\begin{equation}
W_\text{sc}= -\frac{1}{2}\kappa_0^2 \int d^4 v \sqrt{|g(v)|}\,\rho(v)\,\int d^4 v^\prime \sqrt{|g(v^\prime)|}\,\rho(v^\prime)\,D(v,v^\prime)\,.
\label{W}
\end{equation}
For two point like sources moving along the trajectories  $v_i(s_i)$  which are parametrized in terms of their  proper times $s_i$, we find
\begin{equation}
\sqrt{|g(v)|}\rho(v)= \sum_{i=1,2}\int ds_i\, \delta^4(v-v_i(s_i))\,.  
\label{rho}
\end{equation}
We assume  the sources to be at rest in Rindler space and evaluate $W_\text{sc}$ by expressing  the proper times $s_i$ by the   coordinate times $\tau_i$ and obtain for  sources positioned at $\xi_i,{\bf x}_{\perp\,i} $
\begin{equation}
W_{\text{sc}}= -\frac{1}{2} \sum_{i,j=1,2}\kappa(\xi_i)\kappa(\xi_j)\int d\tau_i \int d\tau^\prime_j D\big(\tau_i-\tau_j^\prime, \xi_i,\xi_j, {\bf x}_{i\,\perp}- {\bf x}_{j\,\perp}\big)\,.
\label{inse}
\end{equation}
We have introduced the effective coupling constant
\begin{equation}
\kappa(\xi)= e^{a\xi}\kappa_0\,,
\label{efco}
\end{equation}
which, due to the difference between proper and coordinate times,``runs'' with the coordinates $\xi_i$ of the charges. With the sources at rest, $W_{\text{sc}}$  depends only on the differences of the times $\tau_i$. After carrying out the $\tau_{i}$ integrations,   up to a factor $T$, the size of the interval in the integration over the sum of the times,  $W_{\text{sc}}$ is determined by the static propagator and yields for $i\neq j$ the interaction energy of two scalar sources
\begin{eqnarray}
V_{\text{sc}}&=&-\kappa(\xi_1)\kappa(\xi_2) \tilde{D}\big(\xi_1,\xi_2,{\bf x}_{\perp\,1}-{\bf x}_{\perp\,2}\big)=  -\frac{a\kappa_0^2}{4\pi\sinh \eta}\Big[ 1+\frac{ i \eta}{\pi} \Big],
\label{vsc}
\end{eqnarray}
where $\eta=\eta(\xi_1,\xi_2,{\bf x}_{1\,\perp}-{\bf x}_{2\,\perp})$ (cf.\,Eq.\,(\ref{ueta})).
For two sources at rest in Minkowski space, this procedure yields the  interaction energy $-\kappa_0 ^2/\big|{\bf x_1}-{\bf x}_2\big|$.

The interaction energy (\ref{vsc}) constitutes an explicit example of the bivalent role of the acceleration $a$. The $a$ dependence of the real part of the interaction is exclusively due to the dependence of the Hamiltonian (\ref{hamsca}) on the parameter $a$   while the imaginary part depends  in addition on the acceleration  via the temperature $\beta=2\pi/a$ (cf.\,Eq.\,(\ref{saripo2a})). 
The appearance of a non-trivial  imaginary contribution to the ``static interaction''  generated by exchange of scalar particles  and, as we will see also  by photons or gravitons, is a novel phenomenon not encountered in the static interactions in Minkowski space. Here we will analyze this phenomenon. Other properties of the interaction (\ref{vsc}) will be discussed later in the comparison with the ``electrostatic'' interaction. 

As follows from Eq.\,(\ref{ripo2})  the static propagator (\ref{saripo2a}) satisfies the Poisson equation for a point-like source in Rindler space. Since the source is real, the imaginary part of the propagator satisfies the corresponding (homogeneous) Laplace equation. In turn, this implies that the imaginary part of propagator or the  interaction energy can be represented by  a linear superposition of zero modes of the Laplace operator (\ref{scala}). From Eq.\,(\ref{Dur})  we read off 
\begin{eqnarray}
\text{Im}\, \tilde{D}(0,\xi,\xi^\prime,{\bf x}_\perp-{\bf x}_\perp^\prime) &=&\frac{a\, \eta} {4\pi^2  e^{a(\xi+\xi^\prime)}\sinh \eta}\nonumber\\
&=&\frac{1} {4\pi^3 a}\int d^2  k_\perp e^{i {\bf k}_\perp ({\bf x}_\perp -{\bf x}^\prime_\perp)} K_0\Big(\frac{k_\perp}{a}\, e^{a\xi}\Big)\,K_0\Big(\frac{k_\perp}{a}\, e^{a\xi^\prime}\Big)\,.
\label{isp}
\end{eqnarray}
It is instructive to compare  the Rindler space propagator with the finite temperature propagator in Minkowski space. As above we decompose  the propagator of a non interacting scalar field  into thermal and non-thermal contributions
\begin{eqnarray}
\label{ftmp}
 D_\beta(x) &=&  i\frac{1}{\text{tr}\, e^{-\beta H_M}} \text{tr}\,\Big\{ e^{-\beta H_M} T\big[\phi(x) \phi(x^\prime)\big]\Big\} \nonumber\\ &=&\frac{im}{4\pi^2 \sqrt{-x^2+i\epsilon}}K_1\big(m  \sqrt{-x^2+i\epsilon}\,\big)+ \delta D_\beta(x)\,,
\end{eqnarray}
carry out the Fourier transform of the thermal part
\begin{equation}
\label{ftmp4}
 \delta \tilde{D}_\beta(\omega,{\bf x})= \int_{-\infty} ^\infty dt\,  \delta D_\beta(x) e^{i\omega t}= \frac{i}{2\pi x}\theta(\omega^2 -m^2)  \,\frac{\sin \big(\sqrt{\omega^2-m^2}x\big)}{e^{\beta |\omega|}-1}\,,
\end{equation}
and obtain for  massless particles    
\begin{equation}
\label{ftmp5}
m=0\,,\quad \tilde{D}_\beta(0,{\bf x}-{\bf x}^\prime)= \int_{-\infty} ^\infty dt\,  D_\beta(t,{\bf x}-{\bf x}^\prime)=\frac{1}{4\pi |{\bf x}-{\bf x}^\prime|} + \frac{i}{2\pi \beta}\,.
\end{equation}
The Fourier transformed  thermal propagators in Rindler (\ref{saripo2a}) and in Minkowski space (\ref{ftmp5}) become identical to order $O(a)$ apart from the factor $e^{-a(\xi+\xi^\prime)}$ and differ from the corresponding ground state contribution only by the constant $i/2\pi \beta$. The convergence to this limit is not uniform in Rindler space, since it requires $a|\xi|\,,a|\xi^\prime|\ll 1$.  Due to the twofold role of the acceleration $a$  the finite temperature contribution to the propagator in Minkowski space is only part of the  leading order correction to the propagator (cf.\,Eq.\,(\ref{saripo2a})) in Rindler space.   The structure of the Minkowski space static propagator suggests that the imaginary part is due to  on-shell propagation of zero-energy massless particles. The difference between Rindler and Minkowski space propagators is due to the different dimensions (0 and 2 respectively) of the space  of zero modes.   Furthermore, while  the Minkowski space zero mode is constant, the zero modes in Rindler space exhibit a non-trivial dependence on all the three coordinates. Finally for massive particles no zero mode exists in Minkowski space (Eq.\,(\ref{ftmp4})) while in Rindler space together with the degeneracy in the spectrum also the zero modes persist. In this case the spectral representation in Eq.\,(\ref{isp}) remains valid provided we replace  $k_\perp^2\to k_\perp^2+m^2$ in the arguments of the McDonald functions.  

The imaginary part of the propagator determines the  particle creation and annihilation rates. 
To leading order in the coupling constant $\kappa_0$, the probability for a change in the initial  state in the time interval $[\tau_0,\tau]$  is given by (cf.\,Eqs.\,((\ref{coac})-(\ref{efco})) 
\begin{eqnarray}
&&\hspace{-.6cm}P_{ex}(\tau,\tau_0)= 1-\Big|\langle 0_M| T\, e^{-i\int_{\tau_0}^\tau d\tau^\prime H_{\text{int}}(\tau^\prime)} |0_M\rangle\Big|^2\hspace{-.09cm}\approx\hspace{-.05cm}\text{Re} \int_{\tau_0}^\tau \hspace{-.1cm}d\tau^\prime\hspace{-.1cm}\int_{\tau_0}^\tau\hspace{-.1cm}d\tau^{\prime\prime}\langle  0_M\big|T\big[ H_{\text{int}}(\tau^\prime) H_{\text{int}}(\tau^{\prime\prime})\big]\big| 0_M\rangle\nonumber\\ &&\hspace{-.6cm} \approx\text{Im}\int_{\tau_0}^\tau d\tau^\prime \int_{\tau_0}^\tau d\tau^{\prime\prime}\sum_{i,j=1,2} \kappa(\xi_i) \kappa(\xi_j)  D(\tau^\prime-\tau^{\prime\prime},\xi_i,\,\xi_j, {\bf x}_{\perp\, i}-{\bf x}_{\perp\, j})\,.
\label{px}
\end{eqnarray}
The  rate for a change in the initial state within an arbitrarily  large  time interval $T$ is easily obtained to be 
\begin{eqnarray}
\frac{1}{T}P_{ex}(T,0) &\to& \text{Im} \sum_{i,j=1,2}  \kappa(\xi_i) \kappa(\xi_j) \tilde{D}(0,\xi_i,\,\xi_j, {\bf x}_{\perp\, i}-{\bf x}_{\perp\, j}) \nonumber\\ &=&\frac{\kappa_0^2 a}{2\pi^2} \Big(1+ \frac{\eta(\xi_1,\xi_2,{\bf x}_{\perp\,1}-{\bf x}_{\perp\,2})}{\sinh \eta(\xi_1,\xi_2,{\bf x}_{\perp\,1}-{\bf x}_{\perp\,2})}\Big)
\,.
\label{tora}
\end{eqnarray} 
The total response (not observable in the accelerated frame) of the field to external sources  determines the imaginary part of the static propagator. Furthermore by  defining the reaction rate  in terms of the proper time of the external sources (cf.\,\cite{HIMS921}, \cite{HIMS922}, \cite{CRHM07}, \cite{REWE94})  it is seen that the imaginary part is determined by the  total rate for Bremsstrahlung of uniformly accelerated sources observable in Minkowski space. 

These results imply, that under the transformation from the inertial to the accelerated system comoving with the uniformly accelerated charge,  the particles  generated  in Minkowski space  by Bremsstrahlung are mapped into zero energy excitations in Rindler space. In this way  the conflict between particle production in Minkowski space and the conservation of energy in Rindler space in the presence of static sources is resolved. The on-shell zero modes describe the radiation field in Rindler space without any change in the energy in Rindler space and  may  be viewed as a   ``polarization'' cloud of on-shell particles  induced by external sources or  by  classical detectors (cf.\,\cite{GROV86},\,\cite{MAPB93},\,\cite{UNRU92},\,\cite{HURA00}).  Similar remarks apply for the acceleration induced decay of protons \cite{MULL97},\,\cite{{VAMA00}}. Essential for this important role of the zero modes is, as indicated above, the peculiar symmetry  of the Rindler Hamiltonian which gives rise to the extensive degeneracy.

\section{Wilson and Polyakov loops of the Maxwell field}
\subsection{Wilson loops in Minkowski and Rindler space} 
In this section we consider the Maxwell field coupled to external charges given  by the  action
\begin{equation}
S=-\frac{1}{4}\int d\tau d\xi d^2 x_\perp \sqrt{|g|}  F^{\mu\nu}F_{\mu\nu}+S_{\text{int}}\,,\quad S_{\text{int}}= \int d\tau d\xi d^2 x_\perp \sqrt{|g|} A_\mu\,j^\mu\,.
\label{elact}
\end{equation} 
With changing emphasis we will describe various methods for evaluating  the interaction energy of static sources.  The computation of Wilson loops \cite{BAMU94}  and Polyakov  loop correlation functions \cite{JASM02} constitute  the preferred techniques  in analytical and numerical studies of interaction energies of static sources in gauge theories. In the comparison of these two methods the emphasis will be on  the consequences of the Wick rotation to imaginary time for the interaction energy which  will be seen to  be of relevance also for static interactions in Yang-Mills theories.  Evaluation of the Wilson loops in different gauges will enable us to identify  the  origin  of real and imaginary parts respectively of the electrostatic interaction.

Wilson loops are  defined as  integrals over the gauge field along a closed curve  ${\cal C}$ in space-time
\begin{equation}
e^{i W_{\cal C}} =e^{ie\oint_{\cal C} dx ^\mu A_{\mu}}\,.
\label{WLD}
\end{equation} 
The invariance of the Wilson loop under gauge and  (general) coordinate transformations and reparameterization which is explicit in Eq.\,(\ref{WLD})  makes the Wilson loop a particularly useful tool for our purpose. Up to self energy contributions,  the interaction energy of two oppositely charged sources is given by the expectation value (e.g. in the Minkowski space ground state) of a rectangular Wilson loop in a time-space plane with side lengths $T$ and $R$ 
\begin{equation}
\Sigma_\pm = \lim_{T\to\infty} \frac{1}{T} \widetilde{W}_{{\cal C}[R,T]}\,, 
\label{WLIN}
\end{equation} 
with the ground state expectation value $\widetilde{W}_{{\cal C}[R,T]}$
\begin{equation}
e^{ \widetilde{W}_{{\cal C}[R,T]}}=\langle 0_M|e^{i W_{{\cal C}[R,T]}}|0_M\rangle\,.
\label{wlge}
\end{equation}
The gauge fields along the loop can be interpreted as resulting from two opposite charges which are separated in an initial phase from distance $0$ to $R$ (for a rectangular loop this initial phase is reduced to one point in time), remain  separated at this distance for the  time $T$ and recombine in a final phase. In order to make the contributions from the turning-on period negligible, the interaction energy of static charges is defined by the $T\to \infty$ limit. In terms of the photon propagator 
\begin{equation}
D_{\mu\nu}(x,x^\prime)=  i \langle 0_M\big| T\big[A_\mu(x)\, A_\nu(x^\prime)\big] |0_M\rangle\,,
\label{phpr}
\end{equation} 
the Wilson loop is given by (cf.\,\cite{BAMU94})
\begin{equation}
 \widetilde{W}_{\cal C}=\frac{1}{2}  e^{2} \int ds  \int ds^\prime \,\frac{dx_{\cal C}^\mu}{ds}  \frac{dx^{\prime\,\nu}_{\cal C}}{ds^\prime} D_{\mu\nu}(x_{\cal C}(s),x^\prime_{\cal C}(s^\prime))\,.
\label{pati}
\end{equation}
In Lorenz gauge,  
$$ \partial _{\mu} A^{\mu} = 0,  $$
the Minkowski space photon propagator is expressed in terms of the scalar propagator $D$ (cf.\,Eq.\,(\ref{ripo3})) as 
\begin{eqnarray}
D_{\mu\nu}^M(x,x^\prime)
= \eta_{\mu\nu} D(x,x^\prime ) = \frac{\eta_{\mu\nu}}{4i\pi^2\big[(x-x^\prime)^2 -i\delta\big] }\,,
\label{ripho}
\end{eqnarray}
with  the Minkowski space metric $\eta_{\mu\nu}$.
The Rindler space photon propagator  is obtained by the change in coordinates (\ref{cd3})
\begin{eqnarray}
D_{\mu\nu}(\tau-\tau^\prime,\xi,\xi^\prime,{\bf x}_\perp-{\bf x}_\perp^\prime) &=& \lambda_{\mu\nu} (v,v^\prime)D (x(v), x^\prime(v^\prime))\nonumber\\
&=&\frac{a^2 e^{-a(\xi+\xi^\prime)}}{8i\pi^2 } \frac{\lambda_{\mu\nu} (v,v^\prime)}{\cosh a (\tau-\tau^\prime)  -\cosh \eta -i\delta}\,, 
\label{trpr}
\end{eqnarray}
where we have used the notation 
\begin{equation}
v^{(\prime)}=\{\tau^{(\prime)},\xi^{(\prime)},{\bf x}^{(\prime)}_\perp\}\,,\quad  \lambda_{\mu\nu} (v,v^\prime) =  \frac{dx^\rho}{dv^{\mu}}  \frac{dx^{\prime\sigma}}{dv^{\prime \nu}}\eta_{\rho\sigma} \,.  
\label{rico3}
\end{equation}
Under the coordinate transformation, the Lorenz gauge condition becomes  
$$\nabla_\mu A^\mu = \partial_\tau A^\tau +(\partial_\xi+ 2 a) A^\xi +\boldsymbol{\partial}_\perp {\bf A}^\perp=0\,,$$
with the covariant derivative $\nabla_\mu$.
\subsection{Interaction energy of static charges in Rindler space}
\subsubsection{Wilson loops of gauge fields in Lorenz gauge}
\begin{figure}
\begin{center}
\vskip-2cm
\includegraphics[width=1.\linewidth]{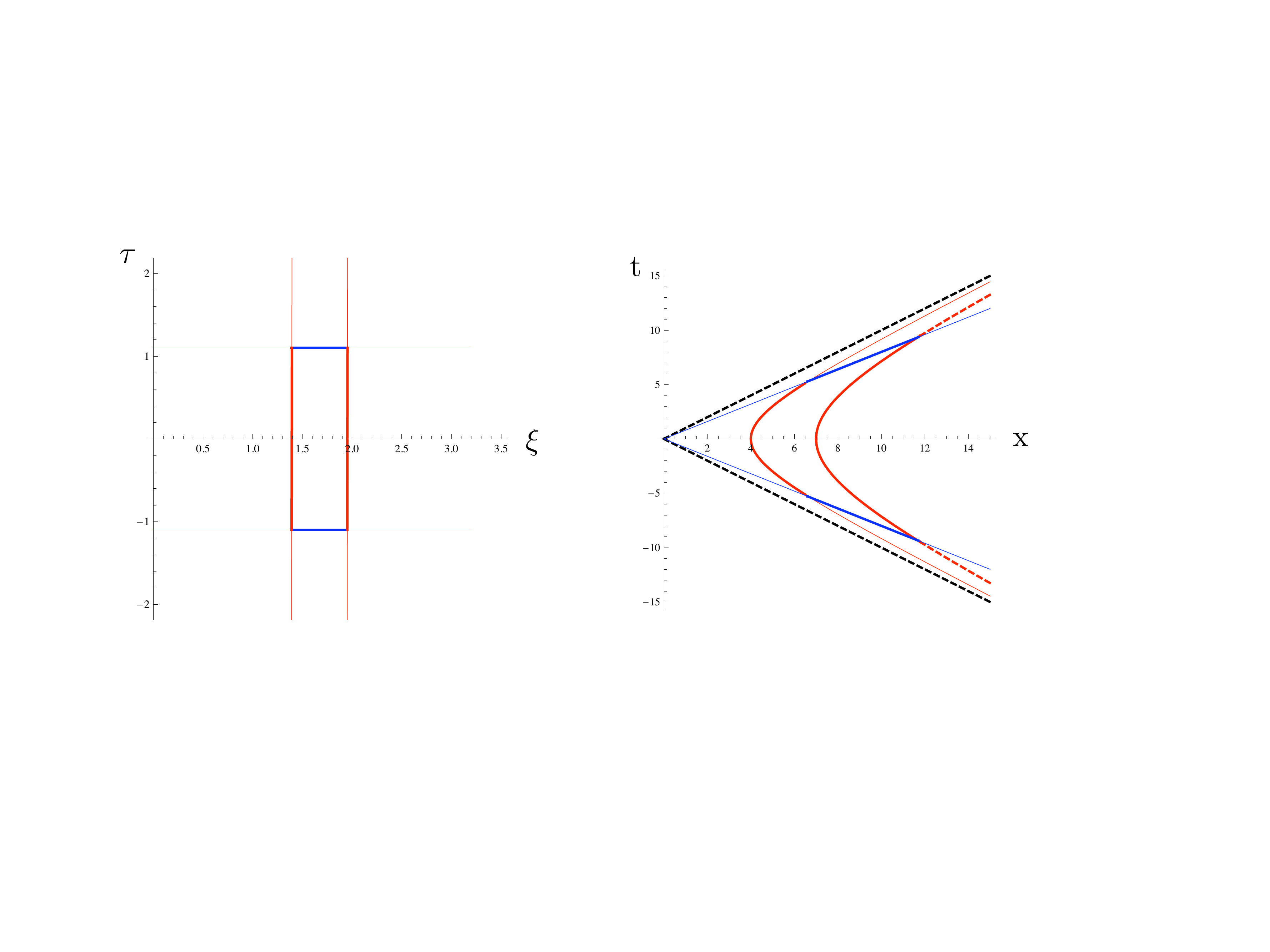}
\vskip-3.5cm
\caption{Wilson loop in Rindler (left) and Minkowski (right) coordinates}
\label{wilmr}
\end{center}
\end{figure}

Invariance of the Wilson loop under coordinate transformations does not imply invariance of the interaction energy. Under the coordinate transformation  (\ref{cd3}) the shape of a  loop changes, as is illustrated in Fig.\,\ref{wilmr} for the case of a rectangular loop in Rindler space.  With the change in shape also  the value of the interaction energy changes which is defined with respect to two different limits ($t \to \infty$ or $\tau \to \infty$). 
We compute the interaction energy for a rectangular loop   with 2 of the 4  segments of the loop varying in time $\tau$  and  $\xi,{\bf x}_\perp$ kept fixed while the other two segments are computed at fixed $\tau$.
The sum $W_{0}$ of the two contributions from the integration along the $\tau$ axis can  be carried out without specifying the segments in the  spatial coordinates. Inserting (cf.\,Eq.\,(\ref{trpr}))  
\begin{equation}
D_{00}(\tau-\tau^\prime,\xi,\xi^\prime,{\bf x}_\perp-{\bf x}_\perp^\prime) =\frac{a^2 }{8i\pi^2} \frac{\cosh a(\tau-\tau^\prime)}{\cosh a (\tau-\tau^\prime)  -\cosh \eta -i\delta}\, 
\label{trpr2}
\end{equation}
into Eq.\, (\ref{pati}) we find
\begin{equation}
\label{W002}
W_{0} = \frac{e^2 a^2}{8 i \pi^2}\int_0^T ds \int_0^T ds^{\prime} \Big[ \frac{\cosh a(s-s^\prime)}{\cosh a(s-s^\prime)-(1+i\delta)}-\frac{\cosh a(s-s^\prime)}{\cosh a(s-s^\prime)-(\cosh \eta +i\delta)}\Big]\,.
\end{equation}
Here    $\cosh \eta$ is given  by (\ref{ueta}) with the spatial coordinates  $\xi^{(\prime)},{\bf x}_\perp^{(\prime)}$ of the vertices of the rectangle. 
Introducing $s\pm s^\prime$ as integration variables $W_0$ can be rewritten as 
\begin{equation}
\label{W003}
W_{0} = \frac{e^2 a^2}{8 i \pi^2}\int_0^T ds \Big[I_0(s,1+i\delta) - I_0(s,\cosh \eta +i\delta)\Big] \,,
\end{equation} 
with  
\begin{eqnarray}
&&\hspace{-1.5cm}I_0(s,\cosh \eta) =2 \int_{0}^{s}ds^\prime \frac{\cosh a s^\prime}{\cosh a s^\prime-\cosh \eta -i\delta}=2s + \frac{2i\pi \cosh \eta}{a\sqrt{\sinh^2\eta +2i\delta}}\nonumber\\
&&\hspace{-1.5cm}\cdot\left[1 -\frac{i}{\pi}\left(\ln \frac{e^{as}-\cosh \eta -\sqrt{\sinh^2 \eta +2 i\delta }}{e^{as}-\cosh \eta + \sqrt{\sinh^2 \eta +2 i\delta }}+\ln \frac{1-\cosh \eta +\sqrt{\sinh^2 \eta +2 i\delta }}{\cosh \eta +\sqrt{\sinh^2 \eta +2 i\delta }-1}\right)\right].
\label{I0}
\end{eqnarray} 
The last step can be verified by differentiation. 
The contribution to the interaction energy $V$ of two oppositely charged sources can be extracted in the $T\to \infty$ limit from the  two $s^\prime$\,-\,independent terms in (\ref{I0}) (the integrals of the $s^\prime$ - dependent terms  converge).

For large $T$,  the  integration along the spatial segments of the rectangle (the horizontal segments of the loop in Fig.\,\ref{wilmr})  yields a $T$-independent  term  and does therefore  not contribute to the interaction energy. The non-diagonal element $D_{01}$ gives rise to two space-time contributions to the Wilson loop which  in the large $T$ limit become independent of the spatial coordinates and  cancel each other. Thus  we obtain  the asymptotic value  $\Sigma$ of the Wilson loop   expressed in terms  $\sigma$ (\ref{ueta})  
\begin{equation}
\Sigma_\pm(\sigma) = \lim_{T\to \infty} \frac{1}{T} W_{0} =  - \frac{e^2}{4\pi}a \coth \eta \big[1+\frac{i\eta}{\pi}\big]+U_{0}= V(\sigma)+U_0\,,
\label{sigma}
\end{equation}
with the interaction energy
\begin{equation}
 V(\sigma)= - \frac{e^2}{4\pi}a \frac{1+\sigma^2}{\sqrt{2\sigma^2+\sigma^4}}\Big[1 +\frac{i}{\pi}\ln \frac{\sqrt{2\sigma^2+\sigma^4}+\sigma^2}{\sqrt{2\sigma^2+\sigma^4}-\sigma^2}\Big] \,.
\label{inen}
\end{equation}
The  integration ``constant'' $U_0$ arises from the first term in (\ref{W002}) and represents the self-energy of the static charges. Regularizing the divergent integrals by point splitting,  $U_0$ is given in terms of the proper distance in AdS$_4$  (cf.\,Eq.\,(\ref{ueta}))
\begin{equation}
U_{0} = \frac{e^2a}{8\pi}\Big(\frac{1}{\sqrt{2}\delta\sigma(\xi)}+\frac{1}{\sqrt{2}\delta \sigma(\xi^\prime)}+\frac{2i}{\pi}\Big)\,,\quad \delta \sigma^2(\xi^{(\prime)})= \frac{a^2}{2}\big(\delta\xi^2+e^{-2a\xi^{(\prime)}}\delta{\bf x}_\perp^2\big).
\label{itc}
\end{equation}
Before discussing the properties of the ``electrostatic'' interaction and the comparison with the static scalar interaction we briefly describe a simple alternative method  for calculating  the interaction energy of two static charges. In analogy with  the corresponding calculation for scalar fields (cf.\,Eqs.\,(\ref{coac})-(\ref{efco}))  the current  $j^\mu$ in Eq.\,(\ref{elact}) generated by two  charges moving along the trajectories  is parametrized as 
\begin{equation}
\sqrt{|g|}j^\mu(v) = \sum_i e_i \int ds_i\frac{dv^\mu_i(s_i)}{ds_i} \delta^4(v-v_i(s_i))\,, 
\label{jmu}
\end{equation}
resulting in the photon-charge vertex
\begin{equation}
S_{\text{int}}= \sum_ie_i\int ds_i A_\mu(v_i(s_i))\frac{dv_i^\mu}{ds_i}\,.
\label{nrv2}
\end{equation}
As for the scalar case (Eqs.\,(\ref{coac})-(\ref{efco})),  the relevant quantity  to be computed is the effective action   which, for charges at rest in Rindler space, yields the sum of interaction and self energies
\begin{eqnarray}
W_{\text{vc}}&=& \frac{1}{2}\sum_{i=1,2}e_ie_j\int ds_i \int ds_j D_{\mu\nu}\big(v_i(s_i),v_j(s_j)\big)\frac{dv^\mu_i(s_i)}{ds_i} \frac{dv^\nu_j(s_j)}{ds_j}\nonumber\\
&=& \frac{1}{2} \sum_{i,j=1,2}e_i e_j\int d\tau_i \int d\tau_j D_{00}\big(\tau_i-\tau_j, \xi_i,\xi_j, {\bf x}_{i\,\perp}- {\bf x}_{j\,\perp}\big)\,,
\label{inseph}
\end{eqnarray}
where the propagators in different coordinates are obtained from each other by the corresponding coordinate transformations (cf.\,Eq.\,(\ref{trpr})). Unlike in the scalar case, for exchange of photons  no factor renormalizes the coupling constants when  changing  from the proper $s_i$ to coordinate time $\tau$. 
We define the  Fourier transform in time of  the $00$-component of the propagator (Eq.\,(\ref{trpr2}))  by the limit 
\begin{eqnarray}
\tilde{D}_{00}(\omega,\xi,\xi^\prime,{\bf x}_\perp)&=& \lim_{T \to \infty}\int_{-T}^{T} d\tau e^{i\omega\tau} D_{00}(\tau,\xi,\xi^\prime,{\bf x}_\perp)\nonumber \\&=& \frac{a^2 \sin \omega T}{4i\pi^2\omega}+ \frac{a}{4\pi}\coth \eta\Big[ e^{-i\frac{\omega\eta}{a}} + \frac{2 i \sin \frac{\omega\eta}{a}}{1-e^{-2\pi\frac{\omega}{a}}}\Big]\,,
\label{ftd00}
\end{eqnarray}
and  disregarding the (divergent) constant, for opposite charges $e=e_1=-e_2$,   the result (\ref{inen})
\begin{equation}
-e^2\tilde{D}_{00}(0,\xi,\xi^\prime,{\bf x}_\perp) = V(\sigma)\,,
\label{D0V}
\end{equation}
is reproduced.
\begin{figure}[ht] \centering
\vskip -.2cm
\includegraphics[width=.5\linewidth]{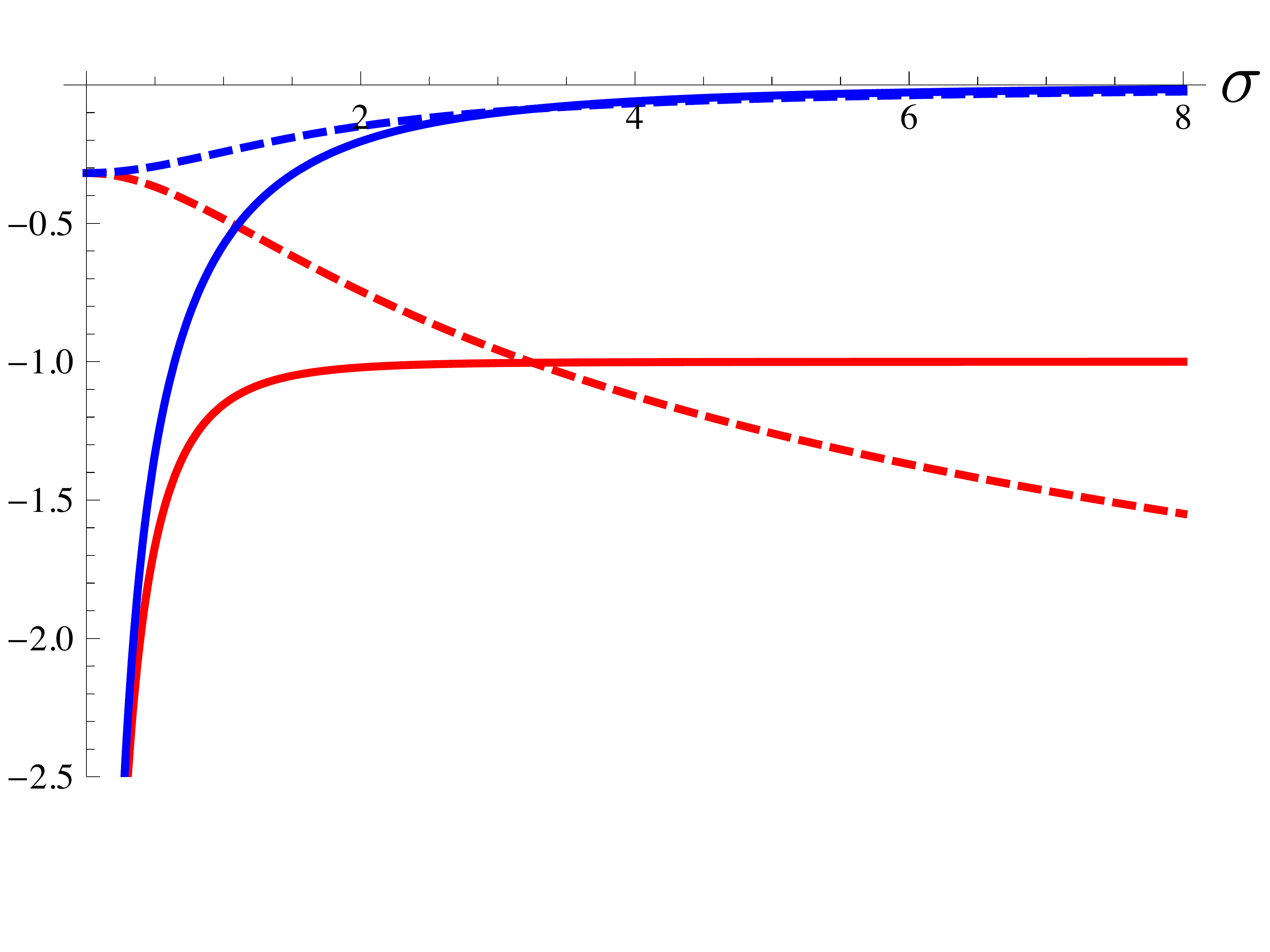}
\vskip-.7cm
\caption{Real (solid) and imaginary (dashed)  parts of the interaction energies $4\pi V_{sc}/a\kappa_0^2$  (\ref{vsc}) and $4\pi V/ae^2$ (\ref{inen}) generated by scalar particle (blue) and photon  (red) exchange respectively as a function of $\sigma$ (\ref{ueta}).}
\label{ripotent}
\end{figure}

In Fig.\,\ref{ripotent} are plotted real and imaginary parts of the interaction energy of two static charges in comparison with the interaction energy of two scalar sources. The distance which determines both scalar  (\ref{vsc}) and  vector (\ref{inen}) interaction energies is  the proper distance of the AdS$_4$ space and not of Rindler space, i.e., the static interaction energies generated by  scalar particle and photon exchange are invariant under the scale transformation (\ref{symm1}).  At small distances where the effect of the inertial force is negligible both interaction energies display the  familiar linear divergence in the real part and as in Minkowski space, vector and scalar exchange give rise to the same behavior. For small distances the imaginary part  is constant  and agrees with the constant in the finite temperature propagator (\ref{ftmp5}) in Minkowski space 
\begin{eqnarray}
\label{ulv}
\lim_{\sigma \to 0} V(\sigma) &=& -\frac{e^2 a}{4\pi}\Big[\frac{1}{\sqrt{2}\sigma}\big(1+\frac{3}{4}\sigma^2\big)+ \frac{i}{\pi}\big(1+\frac{2}{3}\sigma^2\big)\Big]\,,\\
\lim_{\sigma \to 0} V_{\text{sc}}(\sigma) &=& -\frac{\kappa_0^2 a}{4\pi}\Big[\frac{1}{\sqrt{2}\sigma}\big(1-\frac{1}{4}\sigma^2\big)+ \frac{i}{\pi}\big(1+\frac{2}{3}\sigma^2\big)\Big]\,.
\label{ulvsc}
\end{eqnarray}
With increasing distance the inertial forces become important. They weaken both scalar and vector interaction energies  and  change completely  the asymptotic behavior
\begin{eqnarray}
\label{inr}
\lim_{\sigma \to \infty} V(\sigma) &=& -\frac{e^2a}{4\pi}\Big(1+\frac{1}{ 2\sigma^4}+  \frac{i}{\pi} \ln 2\sigma^2\Big), \\
\lim_{\sigma \to \infty} V_{\text{sc}}(\sigma) &=& -\frac{\kappa_0^2a}{4\pi}\frac{1}{\sigma^2}\Big(1+  \frac{i}{\pi} \ln 2\sigma^2\Big)\,.
\label{inrsc}
\end{eqnarray}
Significant differences in the asymptotics for  scalar and vector exchange are obtained.  Qualitative differences are observed for the imaginary part which vanishes asymptotically for scalar exchange and  diverges logarithmically with $\sigma$ for photon exchange. Given  the highly relativistic motion of the sources in Minkowski space, spin effects are expected to be important. Indeed  the comparison of Eqs.\,(\ref{vsc}) and (\ref{sigma}) shows that the differences between scalar and vector exchange  are due to the additional $\cosh \eta$ factor in (\ref{sigma}) arising from the transformation of the propagator from Minkowski to Rindler space. In this context we also note  the stronger asymptotic  suppression of the ``electrostatic'' force in comparison to the force generated by scalar exchange. From the point of view of a Minkowski space observer (cf.\,Eq.\,(\ref{inseph})),  cancellation between magnetic and electric forces generated by the spatial components of the current and by  the charges respectively is to be expected.     

Of particular interest is the behavior of the interaction energy if one of the sources approaches the horizon while the position of the other source is kept fixed.  For
\begin{equation}
\xi_h\to -\infty,\quad \sigma^2 \to \frac{1}{2} e^{-a(\xi_h+\xi)}\big(e^{2a\xi}+a^2({\bf x}_\perp-{\bf x}_\perp^\prime)^2\big)\to\infty\,,
\label{xih}
\end{equation}
the interaction energies are given by
\begin{eqnarray}
\lim_{\xi_h \to -\infty} V(\sigma) &=& -\frac{e^2a}{4\pi}\Big(1-  i\frac{a(\xi_h+\xi)}{\pi}\Big)\,,\\
\lim_{\xi_h \to -\infty} V_{\text{sc}}(\sigma) &=&  -\frac{\kappa_0^2a}{2\pi}e^{a(\xi_h+\xi)}\frac{1}{e^{2a\xi}+a^2({\bf x}_\perp-{\bf x}_\perp^\prime)^2}\Big(1-  i\frac{a(\xi_h+\xi)}{\pi}\Big)\,.
\label{scho}
\end{eqnarray}
In both cases the interaction is dominated by the imaginary radiative contribution and no bound states are expected in this regime.
We also find in analogy with the ``no hair'' theorem  \cite{BEKE72},\,\cite{BEKE98}, \cite{TEIT72} for Schwarzschild black holes that a scalar source close to the horizon cannot be observed asymptotically while vector sources are visible. Formally the difference arises from the running of the scalar coupling (\ref{efco}) to zero when  approaching  the horizon while the electromagnetic coupling remains constant. In detail the results for Rindler and Schwarzschild metrics are different. A significant improvement can be obtained by modifying  the Rindler metric
\begin{equation}
ds^2=e^{2a\xi}(d\tau^2-d\xi^2) -d{\bf x}_\perp^2\to e^{2a\xi}(d\tau^2-d\xi^2) -\frac{1}{4 a^2}\,d\Omega^2\,.
\label{rimo}
\end{equation}
The dynamics in this space is very close to the dynamics in Rindler space unless the difference between the $S^2$ and $R^2$ matters. This is the case when deriving the ``no hair'' theorem where only the $l=0$ waves matter \cite{COWA71}.
\subsubsection{Wilson loops of gauge fields in Weyl gauge}
In the (covariant) Lorenz gauge only the $00$-component of the propagator contributes (cf.\,Eqs.\,(\ref{inen}) and (\ref{D0V})) which simplifies significantly the evaluation of the electrostatic interaction energy and by the same token  hides the difference in origin of its real and imaginary parts. Separation of constrained and dynamical variables becomes manifest in Weyl gauge  
$$A_0=0\,.$$ 
In this gauge one works with two unconstrained dynamical fields  describing the photons and a  longitudinal field constraint  by the Gau\ss\ law. The gauge field is decomposed accordingly 
\begin{equation}
A^i = \partial^i\chi + \hat{A}^i\,,
\label{lotr}
\end{equation}
into the  longitudinal component  given by the scalar field $\chi$ and the transverse field  satisfying  
\begin{equation}
  \label{proj1}
P^{i}_{\;j} \hat{A}^j = \hat{A}^i\,,\quad\text{with}\quad  P^{i}_{\;j} = \delta ^{i}_{\;j} + \partial^{i} \; \frac{1}{\Delta} \;\partial_{j} , 
\end{equation}
and the Laplacian 
\begin{equation}
  \label{lapla}
\Delta = - \partial_{i} \partial^{i}= e^{-2a\xi}\big(\partial_\xi^2 -2a\partial_\xi\big) +\partial^2_\perp \, .
\end{equation} 
Longitudinal and transverse fields are not coupled to each other. Their  actions  are given by
\begin{equation}
S_\ell[\chi]=-\frac{1}{2}\int d^{4}x\partial_0\partial_i\chi\partial_0\partial^i\chi\,,
\label{sl}
\end{equation}
and
\begin{equation}
S_{tr}[\hat{A}]=-\frac{1}{2}\int d^{4}x \partial_0 \hat{A}_i\partial_0 \hat{A}^i-\frac{1}{4}\int d^{4}x \sqrt{|g|}\, g^{ik} g^{jl}
(\partial_{i} \hat{A}_{j} - \partial_{j} \hat{A}_{i}) (\partial_{k} \hat{A}_{l} - \partial_{l} \hat{A}_{k}) \,.
\label{stra}
\end{equation}
The Wilson loop expectation value factorizes into transverse and longitudinal contributions
and  the corresponding interaction energies (\ref{WLIN}) are additive.
In Weyl gauge,  the longitudinal contribution to the Wilson loop for a rectangle with one pair of its sides being parallel to the $\tau$ axis  $(\tau_0=0 \le \tau \le \tau_1= T)$  is given by
$$\oint_{\cal C} dx^i  \partial_i\chi(\tau,\xi,{\bf x}_\perp)= \int_{s_0}^{s_1} ds \frac{dx^i}{ds} \Big(\partial_i\chi(0,\xi(s),{\bf x}_\perp(s))-\partial_i\chi(T,\xi(s),{\bf x}_\perp(s))\Big)= \int d^{4} x\, \rho(x) \chi(x),$$
with 
$$\rho(x) = \sum_{i,j=0,1}(-1)^{i+j+1} \delta(\tau-\tau_i) \delta(\xi-\xi(s_j)) \delta({\bf x}_\perp-{\bf x}_\perp(s_j))\,.$$
In terms  of the Green's function associated with the differential operators $\partial_\tau^2$ and the Laplacian (\ref{lapla})
\begin{equation}
\partial_\tau^2 \frac{1}{2} |\tau| =\delta(\tau)\,,\quad \Delta G(\xi,{\bf x}_\perp,\xi^\prime, {\bf x}_\perp^\prime) = \delta(\xi-\xi^\prime)\delta({\bf x}_\perp-{\bf {x}}_\perp^\prime)\,,
\label{dtg}
\end{equation}
we find the following longitudinal contribution to the Wilson loop
\begin{eqnarray}
W_\ell= -Te^2\big[G(\xi_1,{\bf x}_{\perp\, 1},\xi_0, {\bf x}_{\perp\,0}) -\frac{1}{2}\sum_{i=0,1} G(\xi_i,{\bf x}_{\perp\,i},\xi_i, {\bf x}_{\perp\,i})\big]\,. 
\label{LWL}
\end{eqnarray}
The Green's function $G$  has been calculated in \cite{LOY08}. It coincides with the real part of the static Lorenz gauge  propagator (cf.\,Eq.\,(\ref{ftd00}))
\begin{equation}
G(\xi,{\bf x}_\perp,\xi^\prime, {\bf x}_\perp^\prime) = \text{Re}\, \tilde{D}_{00}(0,\xi,\xi^\prime,{\bf x}_\perp-{\bf x}_\perp^\prime)\,,
\label{glst}
\end{equation}
and therefore up to irrelevant additive constants (cf.\,Eq.\,(\ref{inen}))
\begin{equation}
V(\sigma) = -e^2G(\xi_1,{\bf x}_{\perp\, 1},\xi_0, {\bf x}_{\perp\,0})\,.  
\label{iewg}
\end{equation}

The transverse gauge field operators required for calculating the transverse contribution to the Wilson loop  have been constructed in \cite{LOY08} with the $\xi$ component given by
\begin{equation}
\hat{A}_{1}(\tau,\xi,{\bf x}_{\perp})=e^{2a\xi}\int \frac{d \omega}{\sqrt{2\omega}} \frac{d^{2}k_{\perp}}{2\pi}\frac{k_\perp}{\omega } \Big[a_1 (\omega, {\bf k}_{\perp}) e^{- i \omega \tau + i {\bf k} _{\perp} {\bf x}_{\perp}}
+ \text{h.c.}\Big]\,
\phi_{\omega,{\bf k}_\perp}(\xi,{\bf x_\perp})\, .
\label{notr1a}
\end{equation}
In order to avoid technical complications we assume the Wilson loop to be located in the $\tau$-$\xi$ plane. Applying the general 
expression (\ref{pati})  we obtain
\begin{equation}
W_{tr} (T, \xi_1, \xi_2, {\bf{0}}_{\perp}) = e^2
\int_{\xi_1} ^{\xi_2} d\xi \int_{\xi_1} ^{\xi_2} d\xi'
\{D^W_{11}(0,\xi,\xi',{\bf{0}}_{\perp}) - D^W_{11}(T,\xi,\xi',{\bf{0}}_{\perp})\},
\end{equation}
with  the $11$ component of the Weyl gauge propagator 
\begin{eqnarray}
&&\hspace{-.5cm} D_{11}^W(\tau, \xi, \xi',{\bf x}_{\perp})= i \langle 0_M\big| T\big[\hat{A}_{1}(\tau, \xi, {\bf x}_{\perp})\hat{A}_{1}(0, \xi', {\bf 0}_{\perp})\big]\big|0_M\rangle=-i e^{2a(\xi + \xi')}\int_0 ^{\infty} \frac{d\omega}{\pi^2 a} \sinh{\frac{\pi \omega}{a}}  \nonumber  \\
&&\hspace{-.5cm}\cdot 
\int \frac{d^2 k_{\perp}}{(2\pi)^2} 
\frac{{k_{\perp}}^2}{{\omega}^2} 
(e^{-i\omega |\tau|} + 
\frac{2\cos{\omega \tau}}{e^{\frac{2\pi \omega}{a}} -1}) 
K_{i\frac{\omega}{a}} \Big(\frac{k_{\perp} }{a} e^{a\xi}\Big) K_{i\frac{\omega}{a}}  \Big(\frac{k_{\perp} }{a} e^{a\xi^\prime}\Big) 
e^{i\bf{k}_{\perp} \bf{x}_{\perp}}\,.
\label{D11}
\end{eqnarray}
As above, the interaction energy is determined by the linearly divergent contribution to the Wilson loop  in the large time limit  $aT \gg 1$ and is generated by the thermal contribution in Eq.\,(\ref{D11}).  We obtain for the transverse contribution to the  Wilson loop 
\begin{eqnarray}
W_{tr} (T,\xi_1,\xi_2,{\bf{0}}_{\perp})
=  -iT \frac{e^2a}{4\pi^2} \big(\eta \coth{\eta} - 1\big)\,,
\end{eqnarray}
which  coincides with the imaginary 
part of $V(\sigma)$ (\ref{sigma})  for $\eta= a(\xi_1-\xi_2)$.

With the unambiguous separation of constraint and dynamical degrees of freedom in Weyl gauge, our  result underlines the dynamical origin of the imaginary part of the Wilson loop in the large $T$ limit as opposed to the electrostatic origin of the real part. We essentially can apply now the arguments of Sect. \ref{sec:ipsp} concerning  the imaginary part of the scalar propagator. The photons generated by an accelerated charge in Minkowski space are mapped into zero energy (transverse)  photons in Rindler space. Though of zero energy, these transverse photons carry momentum and contribute therefore in a non-trivial way to the interaction energy. They propagate on the mass shell and their contribution to the interaction energy is therefore purely imaginary. Our derivation of the interaction energy of two static sources in Weyl gauge emphasizes the  classical origin of the real part and the quantum mechanical origin of the imaginary part. Thereby it also confirms the relation of the zero modes in Rindler space generating the imaginary part with the Bremsstrahlung photons of the accelerated charge in Minkowski space.
\subsection{Polyakov loop correlator}
In numerical studies of gauge theories at finite temperature on the lattice an important quantity for characterizing the interaction energy of static charges  is the correlation function of Polyakov loops \cite{JASM02}. These studies are carried out on a Euclidean lattice and  have confirmed the existence of  a transition in Yang-Mills theories from the confining to the deconfined phase.   It would be of great interest to extend these  studies  to Rindler space and to follow the fate of the confined phase when approaching the horizon.  In this paragraph we will calculate the Polyakov loop correlator in Rindler space with imaginary  (Rindler) time
(cf.\,Eq.\,(\ref{wiro}))
which due to the  acceleration is a periodic coordinate.  
Up to a multiplication with $i$, the ``Euclidean''  propagators are obtained from the real time  propagators ((\ref{ripo3}) and (\ref{trpr})) by this change of the time coordinate. In particular the relevant component of the gauge field propagator (\ref{trpr2}) is given by 
\begin{equation}
D_{00}^E(\tau_E-\tau^\prime_E,\xi,\xi^\prime,{\bf x}_\perp-{\bf x}_\perp^\prime)= \frac{a^2}{8\pi^2} \frac{\cos a(\tau_E-\tau^\prime_E) }{\cos a(\tau_E-\tau^\prime_E) -\cosh \eta -i\delta}\,.  
\label{ripoe}
\end{equation} 
The periodicity in imaginary time expresses the similarity  of acceleration and finite temperature. Unlike the temperature, the acceleration $a$ also appears together with the spatial coordinates. The Polyakov loop is defined  by
\begin{equation}
P(e,\xi,{\bf x}_\perp) = \exp \Big\{ i e \oint d\tau_E A^E_0(\tau_E,\xi,{\bf x}_\perp)\Big\}\,,
\label{poly}
\end{equation} 
and the Polyakov loop correlator associated with two static charges $e_{1,2}$ located at $\xi_{1,2}, {\bf x}_{\perp\, 1,2}$ is given by 
\begin{equation}
C_P(\xi_1,\xi_2, {\bf x}_{\perp\,1}, {\bf x}_{\perp\,,2}\big)= \langle 0_M|P(e_1,\xi_1,{\bf x}_{\perp\,1})\, P(e_2,\xi_2,{\bf x}_{\perp\,2})|0_M\rangle\,. 
\label{plc}
\end{equation}
Written as a path integral, this correlation function is easily evaluated with the result
\begin{equation}
C_P(\xi_1,\xi_2, {\bf x}_{\perp\,1}, {\bf x}_{\perp\,2}\big) = e^{-\frac{2\pi}{a}(f_{11}+f_{22} +2 f_{12})}\,,
\label{free}
\end{equation}
where the self and interaction energy contributions to the ``free energy''  $f_{ij}$ are given by (cf.\,Eq.\,(\ref{ueta}))
\begin{eqnarray}
f_{ij} &=& \frac{a e_i e_j }{32\pi^3} \int_0^{2\pi} ds \int_0^{2\pi} ds^\prime \frac{\cos(s-s^\prime)}{\cosh \eta(\xi_i,\xi_j,{\bf x}_{\perp\,i}-{\bf x}_{\perp\,j}) -\cos(s-s^\prime)}\nonumber\\
&=&\frac{a e_i e_j }{8\pi}\big(\coth \eta (\xi_i,\xi_j,{\bf x}_{\perp\,i}-{\bf x}_{\perp\,j}) -1\big)\,.
\label{Fij}
\end{eqnarray}
Regularization by point splitting (cf.\,Eq.\,(\ref{itc})) yields the self-energies 
\begin{equation}
f_{ii} = \frac{e_i^2}{8\pi}\frac{1}{\sqrt{\delta\xi^2+e^{-2a\xi_i}\delta{\bf x}_\perp^2}}\,. 
\label{sen}
\end{equation}
The interaction energy of the  two charges
\begin{equation}
V_P\big(\xi_1,\xi_2, {\bf x}_{\perp\,1}, {\bf x}_{\perp\,2}\big) = \frac{a e_1 e_2 }{4\pi}\Big(\coth \eta\big(\xi_1,\xi_2, {\bf x}_{\perp\,1}-{\bf x}_{\perp\,2}\big) -1\Big)\,,
\label{inenp}
\end{equation}
agrees with the real part of the interaction energy obtained in the calculation of the Wilson loop in Lorenz gauge (\ref{inen})  and with the contribution from the longitudinal degrees of freedom (\ref{iewg}).  
 In agreement with the results of \cite{WALD70} concerning the equivalence of propagators defined in static space-time with 
 either real or imaginary times,  the  Rindler space propagator (cf.\,Eq.\,(\ref{trpr2}))  can be reconstructed given the imaginary time propagator 
 (cf.\,Eq.\,(\ref{ripoe})). However, unlike in Minkowski space, this reconstruction may not work separately for single Fourier components such as  the  static component of the propagators.  Apparently,  the difference between imaginary and real time static propagators of the Maxwell field  is due to the non-trivial (imaginary) contributions to the propagator from zero 
 energy photons. For the case of Yang-Mills theories it remains to be seen  how   this missing information could be gained e.g. by studies of appropriate correlation functions.  
\subsection{Static gravitational  interaction in Rindler space}
To complete our discussion of static forces in Rindler space we sketch the calculation of  the gravitational interaction of two point masses $M_{1,2}$  at rest in Rindler space. We apply the same method as above and start with the graviton propagator in Minkowski space which in harmonic gauge is, in terms of the scalar propagator (\ref{ripo3}), given by  
\begin{equation}
D_{\mu\nu,\rho\sigma}=  \chi_{\mu\nu,\rho\sigma} D(x,x^\prime)\,,\quad \chi_{\mu\nu,\rho\sigma}= \frac{1}{2} \big[\eta_{\mu\rho}\eta_{\nu\sigma}+\eta_{\mu\sigma}\eta_{\nu\rho}-\eta_{\mu\nu}\eta_{\rho\sigma}\big]\,,
\label{grpr1}
\end{equation}
and $\eta_{\mu\nu}$ denotes the Minkowski space metric. 
With the relevant Rindler space tensor component of $\chi$,   
\begin{equation}
\chi^\prime_{00,00}= \frac{1}{2}e^{2a(\xi+\xi^\prime)} \cosh 2a(\tau-\tau^\prime)\,,
\label{chip}
\end{equation}
the corresponding element of the propagator in Rindler space ordered  according to the asymptotics in Rindler time $\tau$ reads
\begin{equation}
D^{(R)}_{00,00}(\tau, \xi, \xi',{\bf x}_{\perp})=\frac{a^2}{4i\pi^2}(\cosh a\tau + \cosh \eta) \, e^{a(\xi+\xi^\prime)}+(2\cosh^2 \eta -1) \, e^{2a(\xi+\xi^\prime)}\,D(\tau, \xi, \xi',{\bf x}_{\perp})\,.
\label{tepr}
\end{equation}
As for scalar (\ref{inse}) and vector exchange (\ref{inseph}) we introduce also for graviton exchange the effective action
\begin{equation}
W_{\text{te}}=  -G\sum_{i=1,2}M_iM_j\int ds_i \int ds_j D_{\mu\nu,\rho\sigma}\big(v_i(s_i),v_j(s_j)\big)\frac{dv^\mu_i(s_i)}{ds_i} \frac{dv^\nu_i(s_i)}{ds_i} \frac{dv_j^{\rho}(s_j)}{ds_j} \frac{dv^\sigma_j(s_j)}{ds_j} \,,
\label{inseph2}
\end{equation}
which for the particles at rest in Rindler space represents their interaction and self-energies. 
As above (cf.\,Eq.\,(\ref{ftd00})), for calculating the static interaction energy,  we define  the divergent $\tau$ integrals by integrating over a large but finite interval $[-T/2,T/2]$ 
and obtain from Eq.\,(\ref{inseph2}), after performing for the convergent terms the limit $T\to \infty$, the following expression for the static gravitational interaction energy
\begin{eqnarray}
\label{gravst}
&&\hspace{-.3cm}V_{\text{gr}}(\xi_1,\xi_2,{\bf x}_{\perp}-{\bf x}^\prime_{\perp})=-G \int  d\xi d^2x_\perp  d\xi^\prime d^2x^\prime _\perp T^{00}_{1}(\xi,{\bf x}_\perp) T^{00}_{2}(\xi^\prime,{\bf x}^\prime_\perp)
\tilde{D}^{(R)}_{00,00}(0,\xi, \xi',{\bf x}_{\perp}-{\bf x}_{\perp}^\prime)\nonumber\\ &&
=- G M_1 M_2 \Big\{\frac{a^2}{2i\pi^2}\Big(\frac{1}{a} \sinh \frac{aT}{2} + \frac{T}{2} \cosh \eta \Big)  +\frac{a}{4\pi} \frac{2\cosh ^2 \eta -1}{\sinh \eta}\Big[ 1+\frac{ i \eta}{\pi} \Big]\Big\}\,,
\end{eqnarray}
where $\eta=\eta(\xi_1,\xi_2,{\bf x}_{\perp\,1}-{\bf x}_{\perp\,2})$ (cf.\,Eq.\,(\ref{ueta})) and we have used 
the  energy momentum tensor density of 2 particles at rest in Rindler space
\begin{equation}
T^{\mu\nu}_{i}(\xi,{\bf x}_\perp) = M_i e^{-a\xi} \delta(\xi-\xi_i)\delta({\bf x}_{\perp}-{\bf x}_{\perp\,i}) \delta_{\mu\,0}\delta_{\nu\,0}\,.
\label{emtr}
\end{equation}
We note that the interaction energies for the three cases considered (Eqs.\,(\ref{vsc}), (\ref{sigma}),\,(\ref{gravst})) depend only on the quantity $\eta$ (\ref{ueta}) or equivalently $\sigma$ and are therefore invariant under the scale transformation  (\ref{symm1}).  For scalar particle and graviton exchange the explicit $\xi,\xi^\prime$ dependent factors  of propagators and vertices (cf.\,(\ref{saripo2a}),\, (\ref{coac}) and (\ref{inseph2})) cancel each other  while both  photon propagator (\ref{trpr2}) and vertex (\ref{nrv2}) depend only on $\eta$ 
  
The terms divergent in the $T\to \infty$ limit are imaginary. They therefore satisfy the wave equation and are to be interpreted as the image of the gravitational radiation of the accelerated particles in Minkowski space. Both the imaginary contribution from the radiation and the real ``static'' gravitational interaction diverge with $\eta$. 
If evaluated in  Minkowski coordinates, the source of these divergences of  $W_{te}$  is not the propagator but  the energy momentum tensor  of uniformly accelerated particles
\begin{eqnarray}
\tilde{T}^{\mu\nu}_{i} &=& M_i \frac{a}{x^1}e^{a\xi_0}\delta({\bf x}_{\perp}-{\bf x}_{\perp\, i})  \delta\Big(x^1-\sqrt{t^2+a^{-2}e^{2a\xi_0}}\Big)\nonumber\\ &&\cdot\Big[x^{1\,2} \delta_{\mu\,0}\delta_{\nu\,0}+ x^1 t(\delta_{\mu\,0}\delta_{\nu\,1}+\delta_{\mu\,1}\delta_{\nu\,0})+t^2 \delta_{\mu\,1}\delta_{\nu\,1}\Big]\,.
\label{emtm2}
\end{eqnarray}
The energy density as well as the  density of the momentum in 1 direction and also  the flux of the 1-component of the momentum in the 1  direction increase for sufficiently large positive or negative times like $|t|$. Thus the exponential increase of  the graviton propagator with $\tau$ and with $\eta$  reflects the increase of the energy and momentum density in Minkowski space. Ultimately  this increase invalidates  the linearization of the Einstein equations and the backreaction on the gravitational field by the increasing   energy-momentum tensor in Minkowski space or equivalently  by the  increasing propagator in Rindler space has to be taken into account. 
 
\section{Hydrogen-like systems and the stability of  matter in Rindler space}
\subsection{Non-relativistic limit of accelerated atoms}
As an application of the formal development we will address in this section the issue of stability of matter under uniform acceleration. The radical change of the dispersion relation between energy and momentum of elementary particles makes it very unlikely that matter as observed in Minkowski space will be formed in an accelerated frame.  We will address this issue in a study 
of  the effect of acceleration on hydrogen-like atoms  consisting of an infinitely heavy, pointlike nucleus 
of charge $Z$ and a single electron. We will  present  evidence that the atomic structure is destroyed by uniform acceleration. 

The acceleration $a$ will be chosen to be small  in comparison to the electron mass $m_e$ in order to keep the electron motion in the  rest frame of the nucleus non-relativistic.  
Furthermore, to simplify the calculation we treat the electron as a scalar particle. Keeping only the coupling between electron and nucleus via $A_0$,  the approximate  Lagrangian is 
\begin{eqnarray}
&&{\cal L}  =  \sqrt{|g|} \; (\partial_{\mu} \phi \partial^{\mu} \phi^{\dagger} - m^{2} \phi\phi^\dagger) 
- i e^{2}\big( \tilde{D}_{00} (0,\xi+\xi_Z,\xi_Z, {\bf x}_{\perp}) +U_0^Z(\xi+\xi_Z,\xi_Z, {\bf x}_{\perp})\big) \nonumber\\
&&\hspace{.55cm}\cdot\;\Big [ \phi^{\dagger} (\xi, {\bf x}_{\perp}) \; \dot{\phi} (\xi, {\bf x}_{\perp}) 
- \dot{\phi}^{\dagger} (\xi, {\bf x}_{\perp})\;  \phi (\xi, {\bf x}_{\perp}) \Big ]\,,
\end{eqnarray}
with the coordinates    $\{\xi_Z\,, {\bf x}_{\perp\,Z}=0\}$ and  $\{\xi+\xi_Z, {\bf x}_{\perp} \}$ denoting the position of the nucleus and the electron respectively   and with $\tilde{D}_{00}$ given in Eqs.\,(\ref{inen}) and (\ref{D0V}). Besides the interaction energy, the Lagrangian also contains the lowest order self energy (cf.\,Eq.\,(\ref{itc}))
\begin{equation}
U_{0}^Z(\xi+\xi_Z,\xi_Z, {\bf x}_{\perp}) = \frac{e^2a}{8\pi}\Big(\frac{1}{\sqrt{2}\delta\sigma(\xi+\xi_Z)}+\frac{Z^2}{\sqrt{2}\delta \sigma(\xi_Z)}+\frac{i  (1+Z^2)}{\pi}\Big)\,.
\label{sfe}
\end{equation}
With the Ansatz 
\begin{displaymath}
\phi (\tau, \xi, {\bf x}_{\perp}) = e^{- i\omega\tau} \varphi (\xi, {\bf x}_{\perp}) \,,
\end{displaymath}
the stationary wave equation reads
\begin{eqnarray}
\Big [ - \omega^{2} - \partial^{2}_{\xi} + (- \partial^{2}_{\perp}  + m^{2})\, e^{2 a( \xi_Z+\xi)}
&+& 2 \omega \,\{Z e^{2} \tilde{D}_{00} (0,\xi+\xi_Z,\xi_Z, {\bf x}_{\perp}) 
\nonumber\\
&+& U_{0}^Z(\xi+\xi_Z, \xi_Z, {\bf x}_{\perp}) \} \Big ] \; \varphi (\xi, {\bf x}_{\perp}) = 0 \,.
\label{stated}
\end{eqnarray}
In the non-relativistic approximation the energy is dominated by the mass term 
\begin{equation}
\omega^2=(m +E)^2 e^{2a\xi_Z} \approx (m^2 +2mE)e^{2a\xi_Z} \,,
\label{gnrl}
\end{equation}
and the Hamiltonian associated with the wave equation (\ref{stated}) is given by (cf.\,Eq.\,(\ref{inen}))
\begin{equation}
H= -\frac{1}{2m}\big(e^{-2a\xi_Z}\partial_\xi^2+e^{2a\xi}\partial_\perp^2\big)+e^{-a\xi_Z}\big(V(\sigma)+U_0^Z\big)+ \frac{m}{2}\big(e^{2a\xi}-1\big)\,, 
\label{hamnr}
\end{equation}
with  $E$  denoting an eigenvalue of $H$.
In comparison to the Hamiltonian in Minkowski space both  potential and kinetic energies are significantly modified.  Only by the requirement of weak acceleration, i.e.\,if $a$ is mall in atomic units,  
\begin{equation}
a \rho_B\ll 1\,\quad\text{with the Bohr radius}\quad  \rho_B= \frac{1}{m Z\alpha}\,,
\label{smxi}
\end{equation}
the connection of $H$ to the Minkowski space Hamiltonian becomes transparent.  
In general, the macroscopic coordinate  of the nucleus $\xi_Z$ is large on this scale. 
In this weak acceleration  limit, we replace $V(\sigma)$ by the leading term in the expansion (\ref{ulv}), rescale the variable $\xi$ and the regulator $\delta \xi$ (cf.\,Eq.\,(\ref{sfe}))
\begin{equation}
z=\xi e^{a\xi_Z}\,,\quad \delta z=\delta \xi e^{a\xi_Z}\,,
\label{reha}
\end{equation}
and introduce the local acceleration (cf.\,\cite{RIND01})
\begin{equation}
\quad a_Z= a e^{-a\xi_Z}\, .
\label{az}
\end{equation}
The resulting Hamiltonian reads
\begin{equation}
H= H_0+W = -\frac{1}{2m}\Delta -\frac{Z\alpha}{\sqrt{z^2+{\bf x}_\perp^2}}+ ma_Z z+ W,
\label{hamg}
\end{equation}
with  (cf.\,Eq.\,(\ref{sfe}))
\begin{equation}
W = e^{-a\xi_Z}\big(i\,\text{Im}V(\sigma)+U_0^Z\big)=\frac{(Z^2+1)e^2}{8\pi\sqrt{\delta z^2+\delta{\bf x}_\perp^2}}+\frac{i a_Ze^2}{8\pi^2}\Big((Z-1)^2-\frac{2\,Z}{3}a_Z^2 (z^2+{\bf x}_\perp^2)\Big)\,.
\label{utd}
\end{equation}
In the weak acceleration limit, the real part of the sum of the electrostatic 
self energies  of the electron and the nucleus is a complex constant independent of $\xi$  and can be dropped. 
\subsection{The imaginary part of the Coulomb interaction}
The  imaginary part of the Hamiltonian (\ref{hamg}) reflects  the presence of on-shell propagating zero energy photons. The leading  term is the self energy of the total charge $Z-1$. 
 It increases linearly with the local acceleration $a_Z$ and  expresses  the instability of the bare source towards decay into source and zero energy photons. Thus  a static charge in Rindler space is accompanied by   on-shell zero energy  photons. This ``cloud'' of photons is nothing else than the image in Rindler space  of the Bremsstrahlung photons generated by the accelerated charge in Minkowski space.

In a system of two oppositely charged sources ($Z=1$)  the imaginary part of self energy and  interaction energy contributions cancel each other for small distances (cf.\,Eq.\,(\ref{utd})). In this limit where the opposite charges  neutralize each other, no Bremsstrahlung is generated  in Minkowski space and no zero energy photons in Rindler space. Beyond this limit and due to the peculiar energy-momentum dispersion relation,  a new phenomenon appears. For values of the  acceleration  of the order of the inverse distance of the two charges,  momentum is transferred by the exchange of zero energy photons. As shown in Fig.\,\ref{ripotent}, the corresponding imaginary part of the interaction energy increases logarithmically with the distance $\sigma$ of the two charges giving rise to a long range (imaginary) force between the static charges. With regard to conceptual issues concerning signatures of the Unruh effect,  a detailed  study of the  consequences of this force for the dynamics of electromagnetically bound systems, e.g., the hydrogen atom,  would be of interest. Simple answers can be obtained only in the limit of small acceleration (Eq.\,(\ref{smxi})). In this limit   perturbation theory applied to $\text{Im} W$ 
yields the width of the hydrogen states induced by the acceleration
$$\Gamma_{n\ell}= \frac{e^2}{6\pi^2}a_Z^3\langle n\ell|\,r^2\,| n\ell\rangle = \frac{e^2}{12\pi^2}a^3_Z\rho_B^2 \big[5n^2+1-3\ell(\ell+1)\big]\,.$$
The inverse of $\Gamma$ is the time it takes  a bare hydrogen atom to get dressed by zero-energy photons.  For this dressing to occur on the atomic time scale, the acceleration $a_Z$ has to be of the order of $m\alpha$ and therefore larger by a factor of $1/\alpha^2$ than typical atomic accelerations. With ``macroscopic'' accelerations such as  the acceleration at the horizon of a black hole,  the time it takes to build up the photon cloud is of the order of the age of the universe.  

Due to the dependence of $\text{Im}\,W$ on the electron coordinate,  initial and final hydrogen states are not necessarily identical but it appears that energy conservation forbids such a transition. On the other hand, considering the corresponding process in Minkowski space,  nothing  forbids for instance  emission of a Bremsstrahlung photon by an  infinitely heavy proton followed by absorption of the (on-shell) photon by the electron and thereby producing an excited state of the hydrogen atom. In Rindler space the on-shell photon is mapped into a zero mode and therefore the excitation of the atom occurs without energy transfer. Thus ground and excited states must be degenerate which  in turn  suggests that the initial state  must be an ionized state. In the following more detailed calculations we will show that this indeed is the case. 
\subsection{Decay of  uniformly accelerated hydrogen-like atoms}
In this section we are discussing the effect of the inertial force in hydrogen  like atoms which are at rest in Rindler space. In particular we will discuss the instability of the atomic states due to ionization by the inertial force described by the term linear in $z$ of the Hamiltonian (\ref{hamg}). For the following  discussion the contribution  $W$  to the Hamiltonian (\ref{hamg}) is irrelevant and will be neglected. The  Hamiltonian $H_0$ is the standard non-relativistic Hamiltonian of hydrogen-like systems complemented by the dipole term $ma_Z z$ which is the small $z$ limit of the inertial force $m^2 e^{2a\xi}$  in the wave-equation (\ref{EQM1}).  As required by the equivalence principle, to leading order in the acceleration $a$,  this additional term appears as an external  homogeneous gravitational field.  

Although of different dynamical origin, the Hamiltonian $H_0$  coincides, after reinterpretation of the constants,  with that of a hydrogen-like system in the presence of an external, constant electric field. For the analysis of the spectrum we employed  the techniques developed for the description of the Stark effect \cite{BESA57},\,\cite{HIOS96}, and carried out the  computation of  the shift in the spectrum  and of the lifetime of the ground state by using parabolic coordinates.
\begin{figure}[ht] \centering
\includegraphics[width=.48\linewidth]{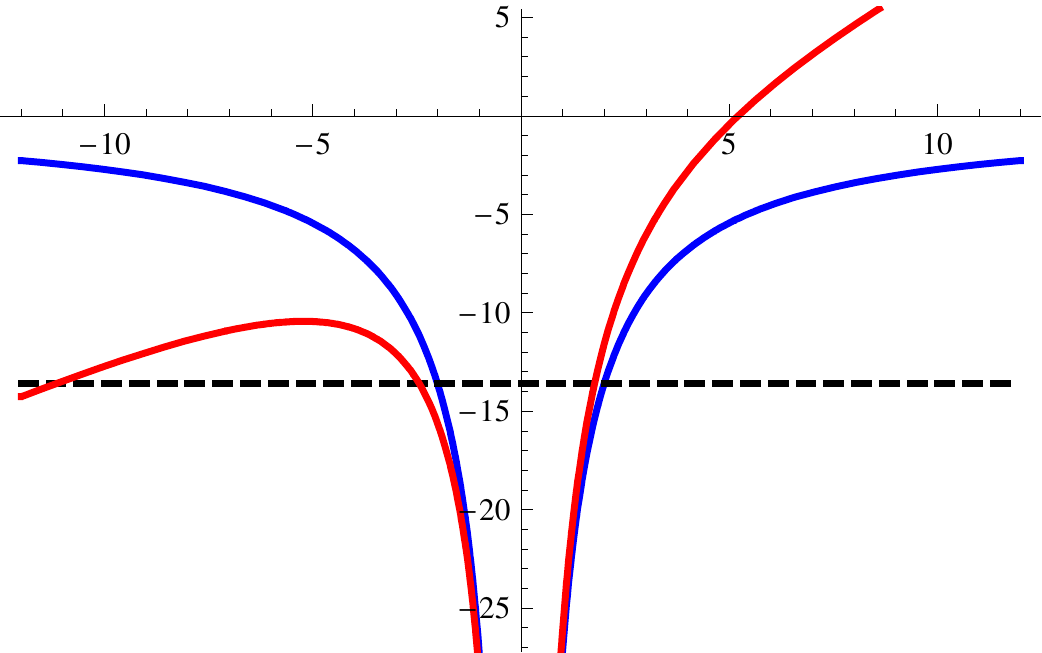}
\caption{Coulomb potential (blue)  and sum of Coulomb and inertial potential (red) in eV along an axis through the origin tilted with respect to the $z$-axis by $60^\circ$ with $a\rho_B\approx 10^{-7}$ with the distance to the origin given in units of the Bohr radius. }
\label{scdr}
\end{figure}
The nature of the spectrum is determined by the interplay between the  inertial force and the electrostatic force  (cf.\,Eq.\,(\ref{inen})) as is demonstrated in Fig.\,\ref{scdr}. The relative strength of inertial and electrostatic interactions 
is given by the parameter
\begin{equation}
\gamma=\frac{a_Z}{m Z^3\alpha^3}\,,
\label{defg}
\end{equation}
which is the ratio between the local acceleration $a_Z$ and the radial acceleration of the electron at the Bohr radius.

When turning on acceleration the first signatures of the presence of inertial forces are changes in the spectrum. The ground state energy of the hydrogen atom is modified in second order perturbation theory  (quadratic Stark effect). For 
$\gamma = 8 \cdot 10^{-3}$, i.e. $a_Z\approx 7\cdot 10^{-20}$\,m\,s$^{-2}$,   the shift $\Delta E$ of the ground state energy $E_0$ is given by
\begin{equation}
\frac{\Delta E}{E_0} = \frac{9}{2}\gamma^2=  0.00029\,.
\label{bacorea}
\end{equation}
Since the leading term vanishes in lowest order perturbation theory, corrections to the Hamiltonian $H_0$ (\ref{hamg}) proportional to $a_Z^2$ have to be considered.  The dominant contribution comes from the expansion of the inertial force in Eq.\,(\ref{hamnr}), $
\Delta_2 H = m a^2_Z z^2$. In comparison to (\ref{bacorea}), the resulting shift of the ground state energy is suppressed by two  powers of $\alpha$  
\begin{equation}
\frac{\Delta_2 E}{\Delta E} = -\frac{4}{9} \alpha^2\,.
\label{dt2}
\end{equation}
Corrections to the Coulomb interaction (cf.\,Eqs.\,(\ref{hamnr}),\,(\ref{ulv})) are suppressed by still higher powers of $\alpha$.

With increasing   acceleration we  also have to account for the decay of the atom induced by the inertial force. As  Fig.\,\ref{scdr} shows,  strictly speaking   no stable hydrogen bound state exists whatever the non vanishing value of the acceleration may be,  thereby confirming the above conjecture that  existence of the imaginary part together with energy conservation in Rindler space is not compatible with the existence of atomic bound states. If on the other hand we assume that the acceleration has not been present for infinite  time i.e. the atom has not been close to a black hole infinitely long,  the lifetime of the atom  is a useful quantity to characterize the effect of the acceleration on the stability of matter. In such a case, the atom decays by tunneling of the electron through the potential barrier.   With decreasing  $\gamma$   the potential energy  approaches the Coulomb potential and the  lifetime of the metastable state tends to infinity. Above  a  critical value  $\gamma=\gamma_c$ where the two turning points  coalesce,  a meaningful definition of a metastable state becomes impossible. In this regime the inertial force is dominant and no bound states are formed. 

Within the formulation in terms of parabolic coordinates the critical  value can be computed  with the results
\begin{equation}
\gamma_c=2(3+\sqrt{7})/(2+\sqrt{7})^3 = 0.11\,,
\label{gc}
\end{equation}
and  the lifetime of the ground state of  hydrogen-like systems can be estimated by   calculating  the tunneling probability in  the WKB approximation.
The resulting expression for the width reads 
\begin{equation}
\Gamma = |E_0| \frac{8}{\gamma} e^{-\frac{2}{3\gamma}}\,.
\label{dera}
\end{equation}
The decay rate of the hydrogen  atom is determined by the tunneling probability 
of the electron $\sim \gamma^{-1}\exp\{-2/3\gamma\} $ and the frequency $\sim |E_0|$ with 
which  the electron reaches the classically forbidden region. With the above choice of $8\cdot 10^{-3}$ for the parameter $\gamma$  the  lifetime of the hydrogen atom is of the 
  order of the age of the universe, while, when  increasing the acceleration by a factor of 14 
  to reach the critical value $\gamma_c$, the width is about $2\%$ of the ground state energy and  no metastable state is formed.

\subsection{Instability of matter}
The decrease of the lifetime  of the hydrogen atom  is a consequence of  the dependence of the acceleration on the position of the nucleus  and is another signature of the spatial variation of the Tolman temperature appearing for instance in the energy momentum tensor of photons in Rindler space \cite{CADE77},\,\cite{SCCD81}. As a consequence, in an accelerated ``rocket'', with  the local acceleration, also the degree of ionization varies within the rocket  as 
   described by $a_Z$ (Eq.\,(\ref{az})). For interpretation of these results 
    in terms of 
    an ensemble of atoms stationary near the horizon of a black hole, it is convenient to introduce the (proper) distance 
    of the nucleus to the horizon
    which, for given $\xi_Z$,  reads (cf.\,Eq.\,(\ref{prdi}),\,(\ref{az}))
\begin{equation}
d_H(\xi_Z) = \frac{1}{a}e^{a\xi_Z}=\frac{1}{a_Z}\,.
\label{dz}
\end{equation}
\begin{figure}[ht] \centering
\includegraphics[width=.48\linewidth]{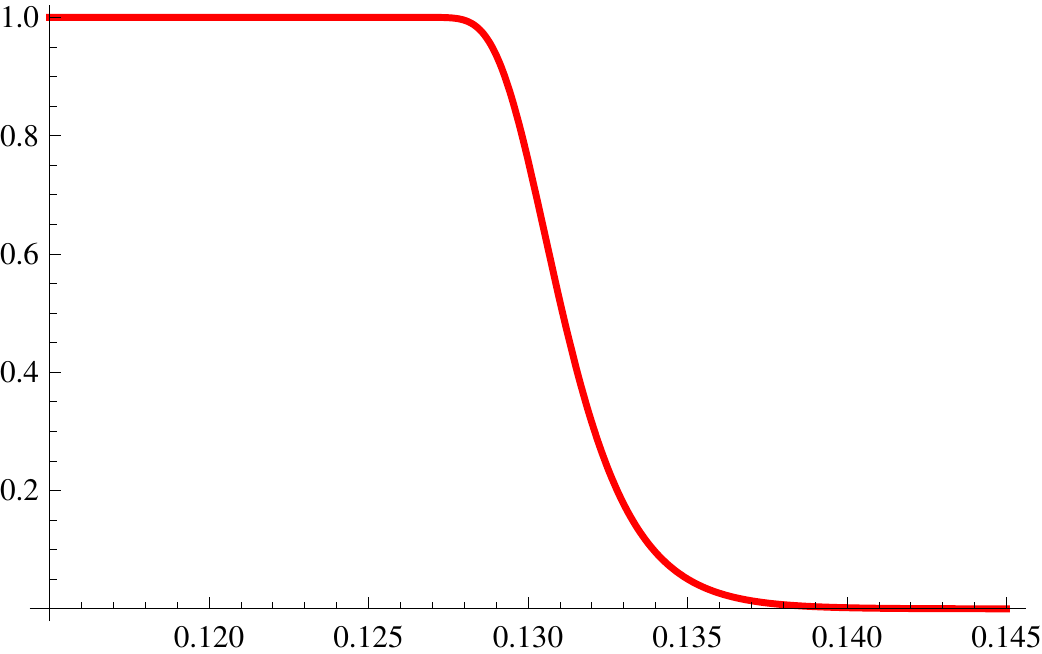}
\caption{Ionization probability of hydrogen atoms as a function of the distance to the horizon (\ref{dz}) of a black hole.}
\label{ionbh}
\end{figure}
In Fig.\,\ref{ionbh} is shown  the probability  that an ensemble of hydrogen atoms has been  ionized since the big bang  as a function of the proper distance  from the horizon e.g. from a Schwarzschild black hole. We note that only the value of the local acceleration $a_Z$ matters (cf.\,Eqs.\,(\ref{defg}),\,(\ref{dera})) and therefore the curve shown in Fig.\,\ref{ionbh} is independent of the acceleration $a$, i.e. of the mass of the black hole. 
The transition from atoms to ions  happens at a distance of about $10^{-4}$\,m within an interval of the order of $10^{-5}$\,m.

In extending these arguments to other forms of matter we find that heavy atoms (cf.\,Eqs.\,(\ref{defg}),\,(\ref{dera}))  are ionized by the inertial force at a distance from the horizon   of the order of  $10^{-10}$\,m. 
Using similar qualitative arguments based on the approximate identity of inertial and the strong force
in the region of ``ionization'',
ejection of a nucleon from the nucleus by acceleration is estimated to occur at a distance to the horizon of the order 
of $10^{-13}$ m.
However, in this case as well as for heavy atoms, the identification of the interaction in accelerated and inertial frames may become problematic. Since we  expect the strong interaction to be  reduced by the inertial force as is the case for electrostatic interactions, these distances to the horizon are most likely underestimated. As for the atoms,  no bound states of nucleons can exist irrespective of the details of the nuclear force provided the acceleration has lasted for an infinitely long time. In turn,  tunneling processes may also be important for the instability of nuclei.   

As far as  the stability of elementary particles are concerned which are subject to  an infinitely long lasting acceleration, related arguments can be put forward which indicate that also in the presence of interactions their properties in Rindler space will remain to be quite different from those in Minkowski space.  For illustration we   consider  a self-interacting scalar field which, depending on the interaction,  may involve mass generation accompanied by spontaneous symmetry breakdown. The Rindler space Hamiltonian of a self interacting scalar field  is given by (cf.\,(\ref{hamsca}))
\begin{equation}H =  \frac{1}{2}\int d\xi d^2x_\perp \big \{  \pi^{2} + (\partial_{\xi}\phi)^{2}
+   e^{2 a \xi} \big ((\boldsymbol{\partial}_{\perp}\phi)^{2}+V(\phi)\big)\,\big \}
\label{sfin}
\end{equation}
with $V$ chosen such that the minimum of $V$ occurs for $\phi=\pm \phi_0\ne 0$.
The expectation value $\phi_0$ of the scalar field generated by the self interaction obviously remains the same after carrying out the coordinate transformation from Minkowski to Rindler space (cf.\,\cite{UNRU84} for a discussion within the path integral approach). The implications of the symmetry breaking in Rindler space are not as obvious as  in Minkowski space.    Symmetry breaking  with mass generation does not leave signatures in the spectrum  of the effective Hamiltonian in Rindler space.  Furthermore as for non-interacting theories,  in the presence of accelerated sources scalar particle production by acceleration in Minkowski space gives rise to on-shell zero energy particles present in Rindler space. A cloud of scalar on-shell particles is present for both massive  and massless particles. On the other hand, we also cannot invoke the restoration of symmetry, when heating the system,  as an argument for a phase transition by increase of the acceleration $a$. A change in $a$ not only changes the temperature but also the Hamiltonian (\ref{sfin}). Thus it is not obvious that a non-vanishing $\phi_0$ also implies a  change from the symmetric to the symmetry broken phase in the whole of Rindler space. Rather it seems plausible that, in analogy  to the transition  of an ensemble of atoms in Fig.\,\ref{ionbh}, locally either the broken or the restored symmetry is realized  depending on the distance from the horizon. To be more specific we consider the standard $\phi^4$ potential and   expand it around the minimum at $\phi=\phi_0$
\begin{equation}
V(\phi)=V(\varphi,\phi_0)=\frac{\lambda}{8}\phi^2(\phi^2-2\phi_0^2) \approx - \frac{\lambda}{8}\phi_0^4 + \frac{m^2}{2}\varphi^2\,,\quad m^2=\lambda \phi^2_0.
\label{p4p}
\end{equation}
Although formally the minimum in the potential energy $V(\phi)$ is attained at $\phi=\phi_0$ independent of the value of $\xi$, it is also plausible that the relevance of this minimum  depends on the value of $\xi$. It has been argued \cite{PADM10} that, for sufficiently small $\xi$, the potential $V(\phi)$  together with the transverse kinetic term can be neglected resulting in a reduction to a non-interacting massless 1+1 dimensional field theory. Here we give a rough estimate of the value of $\xi$ where the effect of the potential $V(\phi)$ becomes negligible  by equating the fluctuations in the energy density including those generated by the transverse motion
\begin{equation}
 \epsilon(\xi)=e^{-2a\xi}\langle 0_M|:{\cal H}(\xi,{\bf x}_\perp):| 0_M\rangle=\frac{e^{-2a\xi}}{2} \langle 0_M|:\pi^{2} + (\partial_{\xi}\varphi)^{2}
+   e^{2 a \xi} \big ((\boldsymbol{\partial}_{\perp}\varphi)^{2}+ m^2 \varphi\big):| 0_M\rangle\,.
\label{endens}
\end{equation}
with the change  in energy density associated with the local breaking of the reflection symmetry $\phi\to -\phi$ 
\begin{equation}
\label{condt}
\epsilon(\xi) \approx\frac{\lambda}{8}\phi_0^4\,.
\end{equation}
The expectation value  $\epsilon(\xi)$ is taken in the Minkowski space  vacuum with the operator  ${\cal H}(\xi,{\bf x}_\perp)$ being normal ordered with respect to the Rindler space vacuum. Inserting the normal mode expansion (\ref{nomosc}) and carrying out the ${\bf k}_\perp$ integration (cf.\,\cite{RG65}) we obtain for the energy density as a function of the distance to the horizon $d_H=e^{a\xi}/a$ (cf.\,Eq.\,(\ref{dz})) 
\begin{eqnarray}
\epsilon(d_H) = \frac{m^2}{4\pi^3d_H^2} \int_0 ^{\infty} d\chi 
e^{-\pi\chi}
\Big\{(\chi^2 + 1) \Big[K_{1+i\chi}(md_H)
K_{1-i\chi} (md_H)
- \big(K_{i\chi} (md_H)\big)^2 \Big]  \nonumber \\
-\frac{1}{2} K_{i\chi}(md_H) \Big[ md_H
K^{\prime} _{i\chi}(md_H)
- K_{i\chi} (md_H) \Big] \Big\},
\label{eden}
\end{eqnarray}  
where $K^\prime_{i\chi}$ denotes the derivative with respect to the argument. Numerical evaluation of this integral  shows that for  $md_H \le 0.2, $   the energy density  agrees within 10\% or less with the $m=0$ limit which can be calculated in closed form 
\begin{equation}
\epsilon(d_H)\approx \frac{11}{480 \pi^2 d_H^4}\,.
\label{epas}
\end{equation}
The energy densities associated with the symmetry breakdown and with  the fluctuations are of the same order of magnitude for the following value of the distance to the horizon
\begin{equation}
d_H\approx \Big(\frac{11}{60\,\pi^2\lambda}\Big)^{\frac{1}{4}} \frac{1}{\phi_0}\,.
\label{dhsb}
\end{equation}
Identifying the parameters $\lambda$ and $\phi_0$ with those   of the Higgs potential of the standard model (with the Higgs mass $100 \,\text{GeV} \le m \le 250\,\text{GeV}$),  Eq.\,(\ref{dhsb}) yields, in qualitative agreement with the numerical evaluation of Eqs.\,(\ref{condt}) and (\ref{eden}),  $d_H\approx  (4\pm 1)\cdot10^{-19}$ m, i.e., we expect that at these or smaller distances  the presence  of a non-vanishing expectation value of  the scalar field will be irrelevant.   
The expression (\ref{epas}) for the variation in the energy density  can also be interpreted as due to the spatial variation of the  Tolman temperature
\begin{equation}
T_{\text{Tolman}}= \frac{1}{2\pi d_H}\,.
\label{tote}
\end{equation} 
Up to numerical factors, the energy density  (\ref{epas}) is that of  a massless scalar field in 3+1 dimensions at the temperature $T=T_{\text{Tolman}}$. The value  $T_{\text{Tolman}}\approx 80$\,GeV  is of the same order of magnitude as the critical temperature of the electroweak phase transition.
\section{Conclusions} 
The focus of our studies of quantum fields in Rindler space has been on the forces  which are generated by exchange of massless bosons and are acting between sources  at rest in Rindler space. We have presented  detailed results for exchange of scalar particles, photons and gravitons. In comparison to Minkowski space, the most striking difference of the interaction energies  is the appearance of a non-trivial imaginary part in Rindler space. Unlike the (classical)  real part of the interaction energy, the imaginary part is of quantum mechanical origin. The difference between real and imaginary contributions can be lucidly illustrated for photon exchange. In Weyl gauge, longitudinal and transverse degrees of freedom are cleanly separated with the longitudinal field determining the real and the transverse photons the imaginary part.  

In contrast to the universal $1/r$ behavior in Minkowski space,  the real part of the interaction energies in Rindler space exhibits significant differences for the three cases considered. Independent of the spin of the exchanged particles is only the behavior of the forces at short distances where  inertial forces are negligible and the leading terms  agree  with the Minkowski space result. For large separations of the sources, the real part of the interaction energy   of scalar sources  is suppressed and the electrostatic interaction energy   approaches a non-vanishing constant while a divergence is encountered for  gravitational sources.  Significant differences  also appear when approaching the horizon  akin to the differences  which, in the context of the ``no-hair'' theorem,  have been obtained close to the horizon of a Schwarzschild black hole \cite{BEKE72},\,\cite{BEKE98},\,\cite{TEIT72}. 

The existence of an imaginary contribution to the interaction energy is a direct consequence of the  peculiar degeneracy of the spectrum of Rindler space particles  and is  therefore of  the same origin as the  one-dimensional density of states appearing in the energy-momentum tensor of scalar or vector fields \cite{CADE77},\,\cite{SCCD81}. The degeneracy  is a consequence of the invariance of the Rindler space Hamiltonian under scale transformations as is the scale invariance of the static interaction energies.  Of particular relevance for the static interactions is the degeneracy of the zero energy sector of the Rindler particles. The degeneracy is characterized by the values of the 2 momentum components transverse to the direction of the acceleration. The imaginary part is generated by emission, on-shell propagation and absorption of zero energy Rindler particles and is directly related to  the (zero energy)  particle creation and annihilation rates in Rindler space. This result, together with the connection of ``radiation'' in Rindler and Minkowski spaces \cite{HIMS921},\,\cite{HIMS922},\,\cite{REWE94},  implies that the imaginary part of the interaction energy in Rindler space is given by the rate of particle production of a uniformly accelerated charge in Minkowski space, measured by the Rindler time. Thus consistency between the radiation observed in Minkowski space and the zero energy radiation  in Rindler space is established by the existence of a non-trivial zero energy sector which  in turn requires  a symmetry which guarantees the degeneracy. Seen from this point of view it may not come as a surprise that the symmetry under appropriately generalized  scale transformations  persists \cite{LOY08} for massive particles and guarantees consistency between Rindler and Minkowski space formulations also in this case. While negligible for small separations of the two sources, the imaginary contribution dominates the force at large distances for scalar particle and photon exchange. As  the real part, also the imaginary part of the gravitational interaction is divergent. The counterpart of this divergence of  the static  graviton propagator in Rindler space is the divergence of the energy momentum tensor of the ``eternally'', uniformly accelerated mass point in Minkowski space. This divergence indicates that the implicit assumption of a negligible backreaction  on the gravitational field is invalid. 

The weakening of the interaction for scalar particle and photon exchange as well as the appearance  of a non-trivial imaginary contribution  point to  instability of matter in Rindler space.  We have confirmed this instability  in a calculation of the rate of ionization of hydrogen atoms at rest as a function of the distance from the horizon or equivalently of the Tolman temperature.  Similar arguments point to the instability of nuclear matter  which will set in at significantly smaller distances to the horizon. In this hierarchy of instability one may also expect, at still smaller distances, instability of the phases of matter determined by the dynamics of quantum fields. Within the context of a self-interacting scalar field we have shown that the strength of the fluctuations increases when approaching the horizon and we have determined the distance where the fluctuations and the mean field generated by symmetry breaking contribute equally to the energy density. At still smaller  distances, most likely  the expectation value of the scalar field becomes irrelevant  for the dynamics. Using  parameters of the standard model, the corresponding Tolman temperature has been found to be of the same order of magnitude as the temperature of the electroweak phase transition. Of  interest in this context is the issue of a possible confinement-deconfinement transition when approaching the horizon. As in Minkowski space, this issue  most likely has to be settled in numerical simulations. The strong force will have to be determined in calculations of the Polyakov loop correlation functions on a lattice in Rindler space with imaginary time. As in the electrostatic case the  interaction energy determined in this way will be a real quantity. Whether confinement in Rindler space implies a vanishing imaginary contribution to the force or the imaginary part can be reconstructed,  on the basis of the numerically determined spectrum and correlation functions,  has to be clarified.

\section*{Acknowledgments}
F.L. is grateful for the support  and the hospitality at  the En'yo Radiation Laboratory and the Hashimoto Mathematical Physics Laboratory of the  Nishina Accelerator Research Center at RIKEN. This work is supported in part by the Grant-in-Aid for Scientific Research from MEXT (No.\,22540302).


\begin{thebibliography}{99}
\bibitem{FULL73} S. A. Fulling,
   {\em Phys. Rev. D} {\bf 7}, (1973), 2850
\bibitem{BOUL75} D.\,G.\,Boulware,  {\em Phys. Rev.\,D}\,{\bf\,11}, (1975), 1404
\bibitem{DAVI75} P. C. W. Davies, {\em J. Phys. A} {\bf 8}, (1975), 609
\bibitem{UNRU76} W. G. Unruh,   {\em Phys. Rev.\,D}{\bf\,14}, (1976), 870
\bibitem{SCCD81} W.\,D.\,Sciama, P.\,Candelas and D.\,Deutsch, {\em Adv.\,in\,Phys.}\,{\bf 30},\,(1981),\,327
\bibitem{HIMS921} A. Higuchi, G. E. A. Matsas and D. Sudarsky, {\em Phys. Rev. D} {\bf 45}, (1992), R3308
\bibitem{HIMS922} A. Higuchi, G. E. A. Matsas and D. Sudarsky,  {\em Phys. Rev. D} {\bf 46}, (1992), 3450
\bibitem{REWE94} H. Ren and E. J. Weinberg, {\em Phys. Rev. D} {\bf 49}, (1994), 6526, [arXiv:hep-th/{\bf 9312038}]
\bibitem{CRHM07} L.\,C.\,B. Crispino, A. Higuchi and G.\,E.\,A. Matsas, {\em Rev. Mod. Phys.} {\bf 80},\,(2008),\,787,   [arXiv:gr-qc/{\bf 0710.5373}] 
\bibitem{UNRU84} W.\,G.\,Unruh and N.\,Weiss, {\em Phys.\,Rev.\,D}\,{\bf\,29}, (1984), 1656
\bibitem{PASS98}  R.\,Passante, {\em Phys.\,Rev.\,A} {\bf 57},\,(1998),\,1590
\bibitem{MULL97} R.\,M\"uller, {\em Phys.\,Rev.\,D}\,{\bf 56},\,(1997),\,953
\bibitem{VAMA00} D.\,A.\,T.\,Vanzella\,and\,G.\,E.\,A.\,\,Matsas,\,{\em Phys.\,Rev.\,D}\,{\bf 63},\,(2000),\,014010,\,[arXiv:hep-th/{\bf 0002010}]
\bibitem{LOY08} F. Lenz,\,K.\,Ohta and  K.\,Yazaki, {\em  Phys. Rev. D}\,{\bf\,78},\,(2008),\,065026, [arXiv:hep-th/{\bf 0803.2001}] 
\bibitem{RIND01} W. Rindler, Relativity, Special, General and Cosmological, Oxford University Press, 2001
\bibitem{TRDA77}W.\,Trost and H.\,Van\,Dam,  {\em Phys.\,Lett. B}  {\bf 71}, (1977), 149 and  {\em Nucl.\,Phys.\,B}  {\bf 152}, (1979), 442 
\bibitem{DOWK78}J.\,S.\,Dowker,  {\em Phys.\,Rev.\,D}{\bf\,18}, (1978), 1856
\bibitem{CHDU78} S.\,M.\,Christensen and M.\,J.\,Duff, {\em Nucl.\,Phys.\,B}  {\bf 146}, (1978), 11
\bibitem{LINE95} B.\,Linet,  [arXiv:gr-qc/{\bf9505033}] 
\bibitem{SVZA08} N.\,F.\,Svaiter ans C.\,A.\,D.\,Zarro, {\em Class.\,Quant.\,Grav.} {\bf 25}, (2008), 095008
\bibitem{CARA76} P.\,Candelas and D.\,J.\,Raine,  {\em J.\,Math.\,Phys.} {\bf 17}, (1976), 2101
\bibitem{RG65}  I.~S. Gradshteyn and I.~M. Ryzhik, Table of Integrals, Series and Products, Academic Press 1965
\bibitem{EMOT53} A. Erdelyi, W. Magnus, F. Oberhettinger and  F. G. Tricomi, Higher Transcendental 
Functions, Vol. I, McGraw Hill 1953
\bibitem{GROV86} P.\,G.\,Grove,  {\em Class.\,Quant.\,Grav.} {\bf 3}, (1986), 801
\bibitem{MAPB93} S. Massar, R. Parentani  and R. Brout,  {\em Class.\,Quant.\,Grav.} {\bf 10}, (1993), 385
\bibitem{UNRU92} W.\,G.\,Unruh,  {\em Phys.\,Rev.\,D}{\bf\,46}, (1992), 3271
\bibitem{HURA00} B.\,L.\,Hu and A.\,Raval, [arXiv:quant/ph {\bf 0012134}]  
\bibitem{BAMU94} J.\,Baez \& J.\,P.\,Muniain, Gauge Fields, Knots and Gravity, World Scientific, 1994
\bibitem{JASM02} J.\,Smit, Introduction to Quantum Fields on a Lattice, Cambridge University Press, 2002 
\bibitem{BEKE72} J.\,D.\,Bekenstein,  {\em Phys.\,Rev.\,Lett.}{\bf\,28}, (1972), 452
\bibitem{BEKE98} 
 J.\,D.\,Bekenstein, Black Holes: Classical Properties, Thermodynamics and Heuristic Quantization, in: Cosmology and Gravitation, Atlanticsciences, M. Novello\,(Ed.) p.\,1
\bibitem{TEIT72} C. Teitelboim, {\em Phys. Rev.\,D}\,{\bf\,5}, (1972), 2951 
\bibitem{COWA71} J.\,M.\,Cohen and R.\,M.\,Wald,  {\em J.\,Math.\,Phys.}{\bf\,12}, (1971), 1845 
\bibitem{WALD70} M.\,Wald, {\em Commun.\,Math.\,Phys.} {\bf 70}, (1979), 221
\bibitem{BESA57} H.\,A.\,Bethe and E.\,E.\,Salpeter, Quantum mechanics of one- and two-electron atoms, Springer Verlag, 1957
\bibitem{HIOS96}K.-I.\,Hiraizumi, Y.\,Y.\,Ohshima and H.\,Suzuki, {\em Phys.\,Lett.}\,A{\bf\,216}, (1996), 117
\bibitem{CADE77} P. Candelas and D. Deutsch, {\em Proc.\,R.\,Soc.\,Lond. A} {\bf 354},\,(1977),\,79
\bibitem{PADM10}
T.\,Padmanabhan, {\em Rep.\,Prog.\,Phys.} {\bf 73}, (2010), 046901 [arXiv:gr-qc/{\bf 0911.5004}] 
\end{thebibliography}
\end{document}